\documentclass{aastex62}
\usepackage{graphicx}

\received{2018 March 18}
\revised{2018 April 23}
\accepted{2018 April 28}
\submitjournal{ApJ}

\shorttitle{Galactic forces rule dynamics of Milky Way dwarf galaxies}
\shortauthors{Hammer et al.}
\begin{document}

\title{Galactic forces rule dynamics of Milky Way dwarf galaxies}

\correspondingauthor{Francois Hammer}
\email{francois.hammer@obspm.fr}

\author{Francois Hammer}
\affil{GEPI, Observatoire de Paris, Universit\'e PSL, CNRS, Place Jules Janssen 92195, Meudon, France}
\author{Yanbin Yang}
\affiliation{GEPI, Observatoire de Paris, Universit\'e PSL, CNRS, Place Jules Janssen 92195, Meudon, France}
\author{Frederic Arenou}
\affiliation{GEPI, Observatoire de Paris, Universit\'e PSL, CNRS, Place Jules Janssen 92195, Meudon, France}
\author{Carine Babusiaux}
\affiliation{GEPI, Observatoire de Paris, Universit\'e PSL, CNRS, Place Jules Janssen 92195, Meudon, France}
\affiliation{UniversitŽ de Grenoble-Alpes, CNRS, IPAG, F-38000 Grenoble, France }
\author{Jianling Wang}
\affiliation{GEPI, Observatoire de Paris, Universit\'e PSL, CNRS, Place Jules Janssen 92195, Meudon, France}
\affiliation{NAOC, Chinese Academy of Sciences, A20 Datun Road, 100012 Beijing, PR China.}
\author{Mathieu Puech}
\affiliation{GEPI, Observatoire de Paris, Universit\'e PSL, CNRS, Place Jules Janssen 92195, Meudon, France}
\author{Hector Flores}
\affiliation{GEPI, Observatoire de Paris, Universit\'e PSL, CNRS, Place Jules Janssen 92195, Meudon, France}

\begin{abstract}

Dwarf galaxies populating the Galactic halo are assumed to host the largest fractions of dark matter, as calculated from their velocity dispersions. Their major axes are preferentially aligned with the Vast Polar Structure (VPOS) that is perpendicular to the Galactic disk, and we find their velocity gradients aligned as well. %This finding results in a probability of random occurrence for the VPOS as low as $\sim$ $10^{-5}$. 
It suggests that tidal forces exerted by the Milky Way are distorting dwarf galaxies. \\
Here we demonstrate on the basis of the impulse approximation that the Galactic gravitational acceleration induces the dwarf line-of-sight velocity dispersion, which is also evidenced by strong dependences between both quantities. Since this result is valid for any dwarf mass value, it implies that dark matter estimates in Milky Way dwarfs cannot be deduced from the product of their radius by the square of their line-of-sight velocity dispersion. This questions the high dark matter fractions reported for these evanescent systems, and the universally adopted total-to-stellar mass relationship in the dwarf regime. It suggests that many dwarfs are at their first passage and are dissolving into the Galactic halo. This gives rise to a promising method to estimate the Milky Way total mass profile at large distances.
\end{abstract}



\section{Introduction} \label{sec:intro}

Milky Way (MW) dwarf spheroidal galaxies (dSphs) are sufficiently nearby for observing their detailed kinematics in the low-mass and very low-mass regime. In the $\Lambda$CDM context and under the assumption of equilibrium, they are strongly dominated by dark matter (DM), which is supported by the amplitude and radial profiles of their velocity dispersions \citep{Strigari2008,Walker2009a,Wolf2010}.\\

It is quite an enigma that dSphs belong to a gigantic structure that is almost perpendicular to the MW disk \citep{Lynden-Bell1976,Kunkel1976}, the so-called Vast Polar Structure (VPOS, \citealt{Pawlowski2014}). Its importance has been underlined by the discovery of similar gigantic structures of dSphs surrounding M31 \citep{Ibata2013} and CenA \citep{Muller2018}, and by the fact that these structures appear to rotate coherently. For example, the latter property cannot be reproduced by successive infall of primordial dwarfs during a Hubble time \citep{Pawlowski2014}, and the discovery of new MW dwarfs only strengthens this conclusion \citep{Pawlowski2015,Pawlowski2018}.\\\\
\newline\\

\section{MW dwarfs are preferentially aligned with the VPOS.} \label{sec:VPOS}
\begin{figure}
\epsscale{0.8}
\plotone{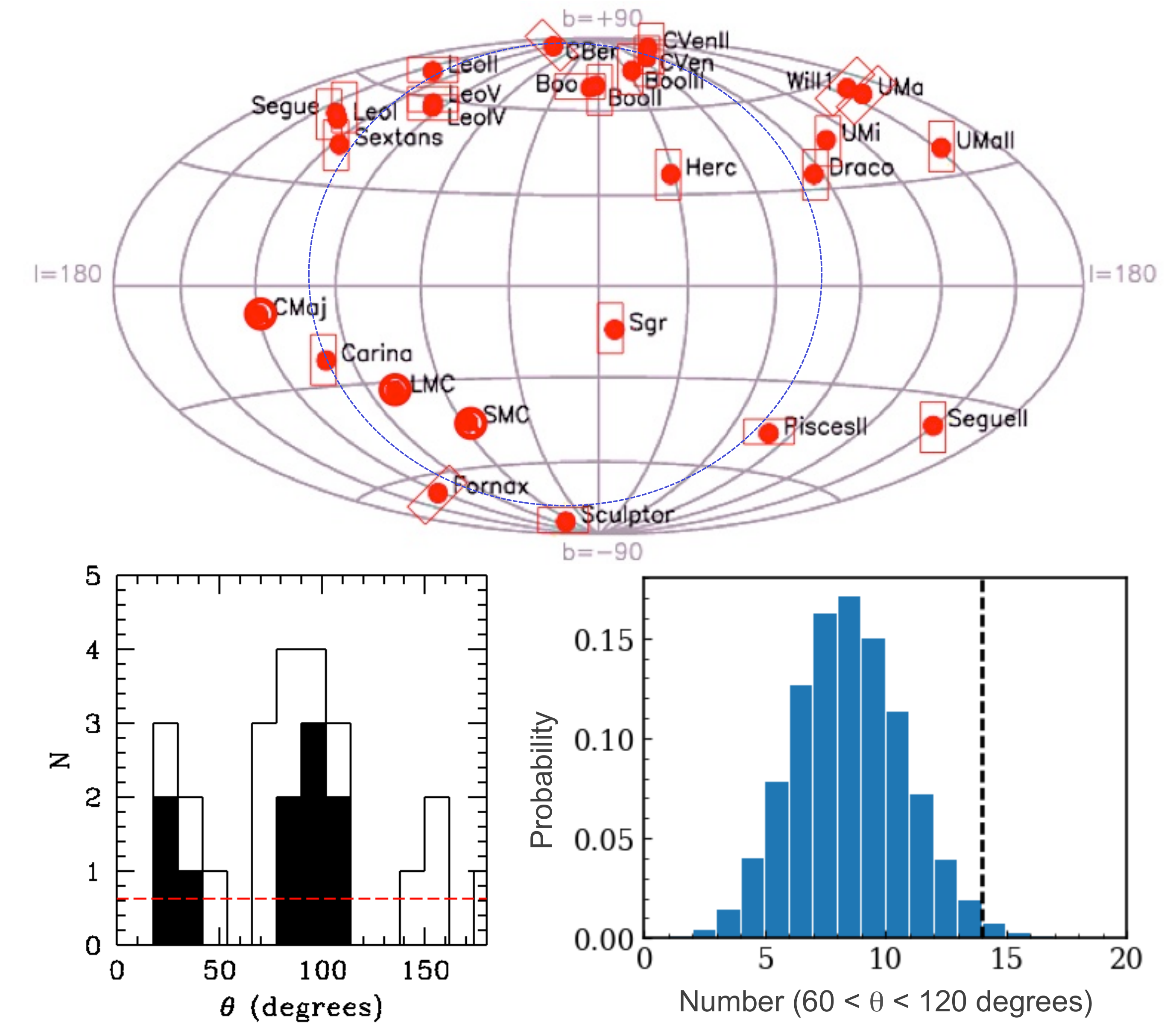}
\caption{{\it(Top)}: distribution of the MW dSphs in Galactic coordinates (l, b) together with the Large and Small Magellanic Clouds and Canis Major. Open rectangles show their major axis orientation $\theta$ as a function of the Galactic plane or latitude, l (horizontal rectangles: $\theta$ $<$ 30 or $\theta$ $>$ 150\degr~; vertical rectangles: 60 $<$ $\theta$ $<$ 120\degr~; 45\degr inclined rectangles: 30 $<$ $\theta$ $< $60\degr~ or 120 $<$ $\theta$ $<$ 150\degr). The blue dotted line indicates the projection of the VPOS, which is seen almost face-on in this projection. {\it(Bottom-left)}: 14 (respectively 7) among 24 (rrespectively 10 of the most massive) dSphs have their major axis with 60 $<$ $\theta$ $<$ 120\degr, respectively. The 10 most massive dSphs are distinguished by the full black histogram. The red-dashed line shows the expectation for a random distribution, revealing the excess of orientations near the VPOS ($\theta$ $\sim$ 90\degr). {\it(Bottom-right)}: result of 100,000 randomized realizations of $\theta$ for a sample of 24 galaxy major axes, the vertical dotted line marking the 14 observed objects having 60 $<$ $\theta$ $<$ 120\degr, with a chance occurrence of $\sim$1\%.
}
\label{Fig1}
\end{figure}

Dwarf spheroidal galaxies are selected from \citet{McConnachie2012} and subsequent updated tables. We only consider secure dSphs to avoid confusion with star clusters, which leads to a sample of 24 dSphs, including the 10 ones with stellar mass larger than $10^5 M_{\odot}$, i.e., by decreasing mass, Sagittarius, Fornax, Leo I, Sculptor, Leo II, Sextans, Carina, Draco, Ursa Minor (UMi) and Canes Venatici.  Top of Figure~\ref{Fig1} evidences an excess of dSphs having their major-axis position angle (PA= $\theta$) within 60 $<$ $\theta$ $<$ 120\degr~ from the Galactic plane or latitude, i.e., aligned with the VPOS ($\theta$ $\sim$ 90\degr). This property had already been identified \citep{Sanders2017}, using a slightly larger number of dSphs that included dSph candidates. The binomial probability that a random distribution of PAs (see the bottom-left panel of Figure~\ref{Fig1}) is consistent with the observations is as low as 0.75\%. We performed 100,000 Monte Carlo simulations with the same number of galaxies while randomizing their PA orientation relative to the VPOS, and Figure~\ref{Fig1} (bottom-right) confirms the low occurrence (1\%) of such an event.  \\

The co-alignment of the dSph PAs led \citet{Sanders2017} to associate this property with the VPOS. This is indeed expected if tidal forces exerted by the MW are distorting the dSphs, in particular if their trajectories are within the VPOS, which seems to be the case for a majority of VPOS dwarfs having proper motion (PM) estimates \citep{Pawlowski2017}. According to \citet{Pawlowski2014} the probability of obtaining a co-rotating VPOS from random distributions is about 0.1\%. Combining this with the fact that most dwarf PAs are aligned along the same direction leads to a chance occurrence for the VPOS as low as $\sim$ $10^{-5}$. \\

Tidal distortion is a common explanation for the Sagittarius dwarf properties, which is further evidenced by the associated gigantic stream \citep{Majewski2003}. Sagittarus is not part of the VPOS although its major axis is also aligned with the Galactic longitude as well as its trajectory. We therefore look at the kinematics of eight dSphs to test the presence of a velocity gradient within the 8 dSphs (see Appendix~\ref{vel_grad}) for which there is a sufficiently large number of stars having radial velocity measurements. We find (see Appendix~\ref{vel_grad}) that MW dwarfs also show a velocity gradient preferentially along the VPOS, which confirms the presence of tidal effects. We notice that no velocity gradient is found within both Leo I and Leo II, presumably because they are at significantly larger distances from the MW than other dwarfs. Let us indeed consider a star located at the half-light radius of Leo I (respectively Leo II): the gravitational acceleration exerted on it by the MW is 32\% (rrespectively 80\%) of that caused by the sole dSph stellar content. Column 8 of Table~\ref{Tab1} gives the ratio of the gravitational acceleration due to the MW to that caused by the dSph stars, and its associated uncertainty.

\section{A strong relation between Galactic acceleration and dynamical-to-stellar Mass.} \label{sec:GAL}

Table~\ref{Tab1} provides all the quantities that have been used throughout this paper. It provides the essential parameters for the 21 dSphs having kinematic measurements and that are in the VPOS. Table~\ref{Tab1} also includes Sagittarius and Crater 2, for comparison purposes. Pisces II and Bootes III are not included in the Table since there is no kinematic available for these galaxies. However, together with the above 21 dSphs and Sagittarius, Pisces II and Bootes III have been used to perform the statistics on the dSph PA orientations in Figure~\ref{Fig1}, which includes a total of 24 galaxies. For completeness, their PA angles from the Galactic plane are $\theta$= -5.0$\pm$5.0 and -64.8$\pm$12, respectively.\\

Values of total masses, stellar masses, line-of-sight velocity dispersions ($\sigma_{los}$) and half-light radii ($r_{half}$) are from \citet{Walker2009a,Walker2010} while distances to the MW center ($D_{MW}$), PA angles ($\theta$), Galactic rest frame velocities ($V_{gsr}$), and other parameters are from \citet{McConnachie2012}. Total (or dynamical), stellar masses, line of sight velocity dispersions ($\sigma_{los}$) and 2D-projected half-light radii ($r_{half}$) values are taken from \citet{Walker2009a,Walker2010}, while distances to the MW ($D_{MW}$) and other parameters are taken from \citet{McConnachie2012}. Total mass and its ratio to stellar mass are taken within $r_{half}$ following an approach \citep{Walker2009a} that leads to a robust and unbiased estimate of the DM content \citep{Strigari2008,Penarrubia2008}. Other quantities (e.g., acceleration ratios, characteristic times) are calculated from the text of this paper.

\begin{longrotatetable}
\begin{deluxetable}{lllrrrrrrlll}
\tablecaption{Essential data for 21 dSphs lying in the VPOS, completed by that for Sagittarius and Crater2 (see text).  Column 1: name; Column 2: distance to the MW; Column 3: 2D-projected half-light radius; Column 4: measured line-of-sight velocity dispersion from \citet{Walker2009a}; Column 5: dSph major-axis PA angle relatively to the MW disk plane; Column 6: total stellar mass evaluated with $M_{\odot}/L_{\odot}$=1 in V-band; Column 7: total-to-stellar mass estimated within $r_{half}$ and that is equal to $(\sigma_{los}/\sigma_{dSph,stars})^2$; Column 8: acceleration ratio as defined by Eq.~\ref{eq1}; Column 9: Galactic rest-frame velocities of dSphs; Column 10: crossing time defined as $r_{half}/\sigma_{los}$; Column 11: ratio of the encounter time ($t_{enc}=D_{MW}/V_{gsr}$)  to the crossing time.\label{Tab1}; Column 12: References for discovery then for distance estimate (see NOTES below).} 
\tablewidth{750pt}
\tabletypesize{\scriptsize}
\tablehead{
\colhead{Name} & \colhead{$D_{MW}$} & 
\colhead{$r_{half}$} & \colhead{$\theta$} & 
\colhead{$\sigma_{los}$} & \colhead{$M_{stellar}$} & 
\colhead{$M_{tot}/M_{stellar}$} & \colhead{$g_{MW}/g_{dSphstar}$} & 
\colhead{$v_{gsr}$} & \colhead{$t_{cross}$} & \colhead{$t_{enc}/t_{cross}$} & \colhead{$ Refs$}\\ 
\colhead{} & \colhead{(kpc)} & \colhead{(kpc)} & \colhead{(degr.)} & 
\colhead{(km s$^{-1}$)} & \colhead{($M_{\odot}$)} & \colhead{(log)} &
\colhead{(log)} & \colhead{(km s$^{-1}$)} & \colhead{$10^{7}$yr} & \colhead{} & \colhead{}
} 
\decimalcolnumbers
\startdata
Segue &   28 & 0.029$\pm$0.007 & 18.3$\pm$8 & 4.3$\pm$1.2 & 330$\pm$210 & 3.275$^{+0.483}_{-0.6835}$  & 3.06$\pm$0.3473 & 117.4 & 0.64 & 35.36 & 1, 2\\
UrsaMaj.II  & 38 & 0.14$\pm$0.025 & -5.9$\pm$4 & 6.7$\pm$1.4 & 4000$\pm$1900 & 3.261$^{+0.3692}_{-0.4697}$  & 3.17$\pm$0.2584 & -35.8 & 1.99 & 50.8 & 3, 4\\
BootesII  & 40 & 0.051$\pm$0.017 & 9.99$\pm$55 & 10.5$\pm$7.4 & 1000$\pm$800 & 3.814$^{+0.7367}_{-6.861}$  & 2.865$\pm$0.4528 & -125.7 & 0.46  & 65.5& 5, 6\\
SegueII  & 41 & 0.034$\pm$0.005 & 23.0$\pm$17 & 3.4$\pm$1.8 & 850$\pm$170 & 2.729$^{+0.4027}_{-5.445}$  & 2.57$\pm$0.1546 & 48.93 & 0.95  & 83.8 & 7\\
Willman1  & 43 & 0.025$\pm$0.006 & 47.41$\pm$5 & 4.3$\pm$1.8 & 1000$\pm$700 & 2.729$^{+0.5764}_{-1.194}$  & 2.204$\pm$0.369 & 31.65 & 0.55 & 233.6 & 8, 9\\
ComaBer.  & 45 & 0.077$\pm$0.01 & 56.6$\pm$10 & 4.6$\pm$0.8 & 3700$\pm$1700 & 2.708$^{+0.3369}_{-0.4013}$  & 2.587$\pm$0.2295 & 76.47 & 1.59 & 35.16 & 1, 4\\
Bootes &   64 & 0.242$\pm$0.021 & 62.97$\pm$6 & 6.5$\pm$2.0 & 30,000$\pm$6000 & 2.597$^{+0.2969}_{-0.5089}$  & 2.462$\pm$0.1151 & 96.46 & 3.54 & 17.82 & 10, 9\\
Draco &   76 & 0.196$\pm$0.012 & 2.49$\pm$2 & 9.1$\pm$1.2 & 0.27$\pm$0.04 $10^6$ & 1.843$^{+0.1685}_{-0.2015}$  & 1.219$\pm$0.0836 & -110.5 & 2.05  & 31.94 & 11, 12\\
UrsaMinor  & 78 & 0.28$\pm$0.015 & 7.84$\pm$5 & 9.5$\pm$1.2 & 0.20$\pm$0.09 $10^6$ & 2.166$^{+0.2954}_{-0.3254}$  & 1.643$\pm$0.2011 & -96.11 & 2.80  & 27.53 & 11, 13 \\
Sculptor  & 86 & 0.26$\pm$0.039 & -64.5$\pm$1 & 9.2$\pm$1.1 & 1.4$\pm$0.6 $10^6$ & 1.261$^{+0.2943}_{-0.3304}$  & 0.6725$\pm$0.227 & 86.56 & 2.69 & 35.16 & 14, 15\\
Sextans(I)  & 89 & 0.682$\pm$0.117 & 5.95$\pm$5 & 7.9$\pm$1.3 & 0.41$\pm$0.19 $10^6$ & 2.081$^{+0.3386}_{-0.4029}$  & 2.022$\pm$0.2507 & 78.58 & 8.21  & 13.12 & 16, 17\\
UrsaMajor  & 102 & 0.318$\pm$0.045 & 49.87$\pm$3 & 11.9$\pm$3.5 & 14,000$\pm$4000 & 3.572$^{+0.3297}_{-0.5276}$  & 2.739$\pm$0.1749 & -10.24 & 2.54 & 372.7 & 18, 4\\
Carina   & 107 & 0.241$\pm$0.023 & -8.85$\pm$5 & 6.6$\pm$1 .2 & 0.24$\pm$0.10 $10^6$ & 1.705$^{+0.3198}_{-0.3859}$  & 1.233$\pm$0.1993 & 22.24 & 3.47 & 131.7 & 19, 17\\
Hercules  & 126 & 0.33$\pm$0.063 & 12.47$\pm$4 & 3.7$\pm$0.9 & 36,000$\pm$11,000 & 2.163$^{+0.3154}_{-0.454}$  & 2.224$\pm$0.2126 & 129.7 & 8.48  & 10.89 & 1, 17\\
Fornax   & 149 & 0.668$\pm$0.034 & 59.06$\pm$1 & 11.7$\pm$0.9 & 14.0$\pm$4.0 $10^6$ & 0.879$^{+0.1894}_{-0.201}$  & 0.1357$\pm$0.132 & -20.71 & 5.43 & 126 & 21, 17\\
LeoIV   & 155 & 0.116$\pm$0.03 & 88.2$\pm$9 & 3.3$\pm$1.7 & 8700$\pm$4600 & 2.226$^{+0.5442}_{-6.095}$  & 1.795$\pm$0.3216 & 14.34 & 3.34 & 307.6 & 1, 4\\
CanesV.II  & 161 & 0.074$\pm$0.012 & 13.09$\pm$9 & 4.6$\pm$1.0 & 7900$\pm$3600 & 2.362$^{+0.5442}_{-6.095}$  & 1.421$\pm$0.2432 & -103.3 & 1.53  & 96.9 & 22, 8\\
LeoV   & 179 & 0.042$\pm$0.005 & 60.2$\pm$13 & 2.4$\pm$1.9 & 4500$\pm$2600 & 1.795$^{+0.6641}_{-5.398}$  & 1.102$\pm$0.2717 & 59.5 & 1.66  & 171.9 & 23, 23\\
CanesV. & 218 & 0.564$\pm$0.036 & 16.73$\pm$4 & 7.6$\pm$0.4 & 0.23$\pm$0.03 $10^6$ & 2.216$^{+0.1071}_{-0.1138}$  & 1.515$\pm$0.0793 & 68.41 & 7.05  & 42.94 & 24, 8\\
LeoII   & 236 & 0.151$\pm$0.017 & 60.4$\pm$10 & 6.6$\pm$0.7 & 0.59$\pm$0.18 $10^6$ & 1.112$^{+0.2261}_{-0.2519}$  & -0.0935$\pm$0.165 & 22.97 & 2.17 & 449 & 25, 12\\
LeoI  & 258 & 0.246$\pm$0.019 & 16.45$\pm$3 & 9.2$\pm$1.4 & 3.4$\pm$1.1 $10^6$ & 0.851$^{+0.2593}_{-0.3043}$  & 0.4922$\pm$0.156 & 178.9 & 2.54  & 53.94 & 25, 26\\
\hline
Sagittarius  & 18 & 1.55$\pm$0.05 &  -11.4$\pm$2 & 11.4$\pm$0.7 & 17.0$\pm$3.0 $10^6$ & 1.138$^{+0.1286}_{-0.1357}$  & 2.052$\pm$0.0817 & 160.8 & 129  & 0.823 & 27, 28\\
Crater2  & 117 & 1.066$\pm$0.084 & -   & 2.7$\pm$0.3 & 0.17$\pm$0.03 $10^6$ & 1.735$^{+0.1705}_{-0.1954}$  & 2.624$\pm$0.1043 & -71.79 & 375  & 4.146 & 29, 29\\
\enddata
\tablecomments{ 1- \citet{Belokurov2007}, 2- \citet{Simon2011}, 3- \citet{Zucker2006a}, 4- \citet{Simon2007}, 5- \citet{Walsh2007}, 6- \citet{Koch2009}, 7- \citet{Belokurov2009}, 8- \citet{Willman2005a}, 9- \citet{Martin2007}, 10- \citet{Belokurov2006}, 11- \citet{Wilson1955}, 12- \citet{Walker2007}, 13- \citet{Walker2009a}, 14- \citet{Shapley1938a}, 15- \citet{Pietrzynski2008}, 16- \citet{Irwin1990}, 17- \citet{Walker2009b}, 18- \citet{Willman2005b}, 19- \citet{Cannon1977}, 20- \citet{Aden2009}, 21- \citet{Shapley1938b}, 22- \citet{Sakamoto2006}, 23- \citet{Belokurov2008}, 24- \citet{Zucker2006b}, 25- \citet{Harrington1950}, 26- \citet{Mateo2008},27- \citet{Ibata1994}, 28- \citet{Monaco2004}, 29- \citet{Caldwell2017}}
\end{deluxetable}
\end{longrotatetable}

\begin{figure*}
\epsscale{1.0}
\plotone{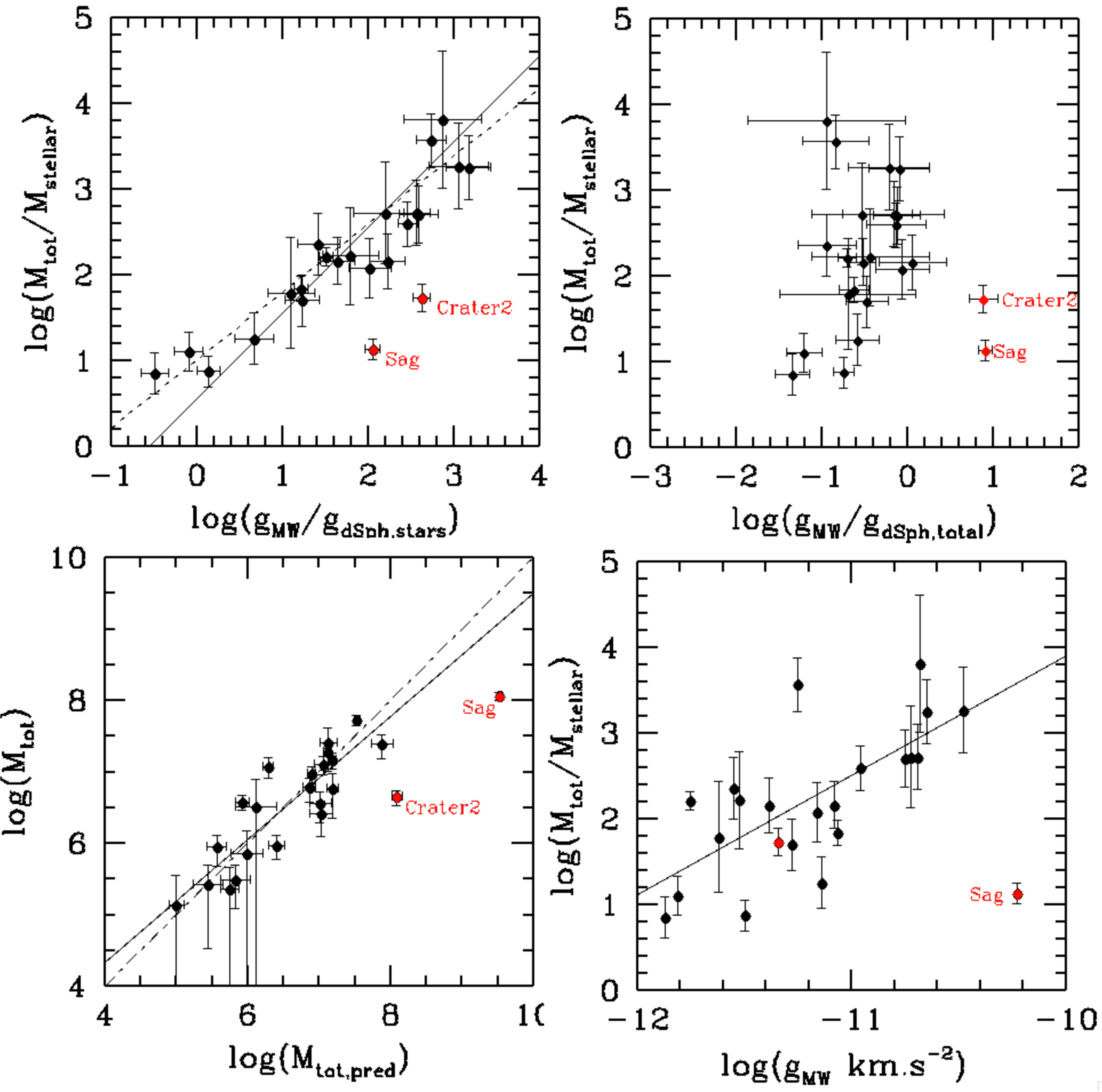}
\caption{Black points represent the dSphs of the VPOS, while the two red points represent Sagittarius and Crater2. Notice that for 3 of the 4 panels the ordinate quantity $M_{tot}/M_{stellar}$ could be replaced by $(\sigma_{los}/\sigma_{dSph,stars})^2$. {\it(Top-left)}: Total-to-stellar mass ratio as a function of the ratio of the MW to dSph acceleration, all quantities being estimated at $r_{half}$. The dashed line shows the strong correlation between the two quantities, while the full line stands for the minimized $\chi^2$ line with a slope equals to 1. The latter leads to a coefficient in Eq.~\ref{eq5} of 3.47 instead of 3.529. {\it(Top-right)}: Same but after replacing the stellar mass in the abscissa by the total mass calculated from \citet{Walker2009a}. {\it(Bottom-left)}: Total mass versus that predicted from the Galactic acceleration (Eq.~\ref{eq5}), with the equality line (short dash - long dash line) matching perfectly the data. The solid line shows the best fit relation to the data. {\it(Bottom-right)}: Total-to-stellar mass ratio as a function of the Galactic acceleration, with the solid line representing the best fit to the data.
}
\label{Fig2}
\end{figure*}

Let us now specify the calculation of the ratio of the MW gravity (or acceleration) to the self-gravity (or acceleration) due to the dSph stellar mass at $r_{half}$ ($M_{stellar}$/2), which is:

\begin{equation}
\frac{g_{MW}}{g_{dSph,stars}}=
\frac{2 M_{MW} ( D_{MW} ) }{M_{stellar}}  \times   (\frac{r_{half}}{D_{MW}} )^{2}  ,
\label{eq1}
\end{equation}

in which, $M_{MW}$($D_{MW}$) refers to the Galactic mass enclosed within a radius $D_{MW}$. We have adopted the mass profile from Eq. 17 and Table 2 of \citet{Sofue2012}, which reproduces the MW rotation curve out to at least 100 kpc. \\
Examination of Table~\ref{Tab1} suggests that galaxies with large dark matter content (ultra-faint dwarfs) have large $g_{MW}$/$g_{dSph}$ values and vice versa. Figure~\ref{Fig2} (upper left panel) reveals a very strong correlation\footnote{Throughout the text we have used a Spearman's rank correlation $\rho$ that doesn't assume any shape for the relationship between variables;  t is distributed as Student's t distribution with n - 2 degrees of freedom under the null hypothesis.} between these two quantities, with $\rho$= 0.93, t= 10.8, resulting in a probability of random occurrence P$< 10^{-10}$ for 19 degrees of freedom. This very tight correlation over more than three decades means that knowing the acceleration ratio (Eq.~\ref{eq1}), one may deduce the total-to-stellar mass ratio with very high accuracy. It remains similar if one assumes a constant DM-to-stellar mass ratio for the dSphs. The correlation significance merely decreases (to $\rho$= 0.9, t= 8.9) if one assumes that the whole MW is a point mass. The upper right panel of Figure~\ref{Fig2} shows that the correlation vanishes ($\rho$= 0.27, t= 1.2 and P= 0.11) when replacing stellar masses ($M_{stellar}$) by total masses in the acceleration ratio (Eq.~\ref{eq1}). The above results point toward a strong link between DM estimates and the MW gravitational acceleration, $g_{MW}$. \\

One may wonder whether such a strong correlation could result from the fact that stellar masses are involved in the two correlated quantities. The correlation revealed by the bottom-left panel of Figure~\ref{Fig2} is still very tight ($\rho$= 0.87, t= 7.65 and P= 1.2 $10^{-7}$) after removing the stellar mass. The bottom-right panel of Figure~\ref{Fig2} evidences that the DM fraction is directly correlated ($\rho$= 0.7, t= 4.3 and P= 1.7 $10^{-4}$) with $g_{MW}$. This questions the hypothesis of neglecting the impact of Galactic forces when evaluating the DM content enclosed within the half-light radius of the dSphs. Hereafter we propose a physical interpretation for these new relations (see also Appendix~\ref{calc} for the details of the calculations).

\section{Galactic forces drive the kinematics and dark-to-stellar Mass ratios of dSphs.} \label{sec:GALTIDE}

In principle the Galactic tidal acceleration should depend on $D_{MW}^{-3}$, but we found (see Appendix~\ref{calc_tidal}) a weaker correlation when replacing in Fig. 2 (top-left panel) $g_{MW}$ by the MW tidal acceleration, $g_{MW,tides}$. After several orbital periods, stellar systems are fully captured by the MW, and their stellar motions are likely dominated by Galactic tides. However, the Magellanic Clouds are known to be at their first passage, as suggested by their proper motions \citep{Kallivayalil2013} and also by their high gas content. Since the Clouds are within the VPOS, one may investigate a scenario in which the VPOS is made of Clouds and dSphs orbiting together, just before and after their first pericenter passage relative to the MW, respectively. \\

Let us consider that the dSph internal radii are small compared to the distance ($D_{MW}$) to the MW  and let us assume that the line of sight are parallel to the directions of the force exerted by the MW on the dSphs stars. According to \citet{Walker2009a}, stars at projected R= $r_{half}$ are selected within a circular annulus, whose corresponding volume is a tube (see Figure~\ref{FigS2} in Appendix~\ref{calc_rhalf}) elongated along the line-of-sight direction. Following \citet{Walker2009a} we also suppose that the dSph stars are distributed into Plummer spheres. We have calculated the difference between the MW potential ($\phi$=-G$M_{MW}$/$D_{MW}$) associated with the two halves of the tube that include the closest and farthest stars relative to the MW, respectively. This leads to (see the detailed calculation in Appendix~\ref{calc_rhalf}):

\begin{equation}
   \Delta  \phi   \approx  \frac{r_{half}}{\sqrt[]{2}}  \times  \frac{GM_{MW}(D_{MW}) }{D_{MW}^{2}}   
\label{eq2}
\end{equation}

in which the MW mass is assumed to be constant over the dSph volumes. We have verified that adopting different density profiles would only affect the scaling factor $2^{-1/2}$ in Eq.~\ref{eq2} by less than a few percent (see Appendix~\ref{calc_rhalf}). Because the encounter velocity (few 100 $kms^{-1}$) is much larger than the star velocities ($\sim$10 $kms^{-1}$), one can consider at first the impulse approximation to be valid (see, e.g., \citealt{Binney1987}) and hence we can neglect the internal motions. Energy conservation leads to an increase of the specific kinetic energy K as:

\begin{equation}
K = \frac{1}{2} <\Delta v^{2}>  \approx  \frac{1}{2} \sigma_{los}^{2}  
\label{eq3}
\end{equation}

Because all the induced velocity vectors ($v$) are parallel to the MW acceleration vector and then to the line of sight, their average $<\Delta$$v^2>$ between 
the two half tubes defined above can be identified to the square of the measured line-of-sight velocity dispersion, $\sigma_{los}^2$. The total mass within 
$r_{half}$, is assumed \citep{Walker2009a} to be $M_{tot}$($r_{half}$) = $\mu$ $r_{half}$ $\sigma_{los}^2$ where $\mu$ = 580 $M_{\odot} pc^{-1} km^{-2} s^{2}$. The almost 
one-to-one correlation seen in Figure~\ref{Fig2} between $M_{tot}$/$M_{stellar}$ and $g_{MW}$/$g_{dSph,stars}$ is therefore also a correlation between 
the $(\sigma_{los}/\sigma_{dSph,stars})^2$ and $g_{MW}$/$g_{dSph,stars}$ ratios, where $\sigma_{dSph,stars}$ = $(M_{stellar}/(2\mu r_{half}))^{1/2}$
 is the line-of-sight velocity dispersion associated to half the dSph stellar mass. This suggests that the excess of kinetic energy found in dSphs is indeed related to the Galactic acceleration; by assuming total energy conservation, $\Delta$$\phi$ = K, we find:

\begin{equation}
 \sigma_{los,MW}^{2}=\frac{\sqrt[]{2 }G M_{MW}(D_{MW})   \times  r_{half}}{ D_{MW}^{2}} = \sqrt{2} \:  g_{MW} \: r_{half} \:  
\label{eq4}
\end{equation}

One can then calculate the total dSph masses at $r_{half}$ predicted if their internal kinematics are dominated by the Galactic acceleration, i.e., $M_{tot,pred}$($r_{half}$)= $\mu$ $r_{half}$ $\sigma_{los,MW}^2$, which are:

\begin{equation}
 M_{tot,pred} \approx  3.529  \times  M_{MW}(D_{MW})   \times   \left( \frac{r_{half}}{D_{MW}} \right) ^{2}  \:  
 \label{eq5}
\end{equation}

The bottom-left panel of Figure~\ref{Fig2} shows that the masses predicted by Eq. 5 precisely match the total, DM-dominated mass derived from \citet{Walker2009a}. Does this imply that estimates of DM in MW dSphs are falsified and that Galactic forces drive their internal kinematics? The top-left panel of Figure~\ref{Fig3} shows that a quadratic combination of $\sigma_{los,MW}$ with the velocity dispersions expected from the stellar masses can predict the measured $\sigma_{los}$ within 1-2 standard deviations.  They correlate reasonably with $\rho$= 0.57, t= 3.0 and P= 3.5  $10^{-3}$, and one may attribute the lower degree of correlation to the fact that the $\sigma_{los}$ values are ranging within a rather modest factor of four.

\section{Discussion}

\begin{figure*}
\epsscale{0.95}
\plotone{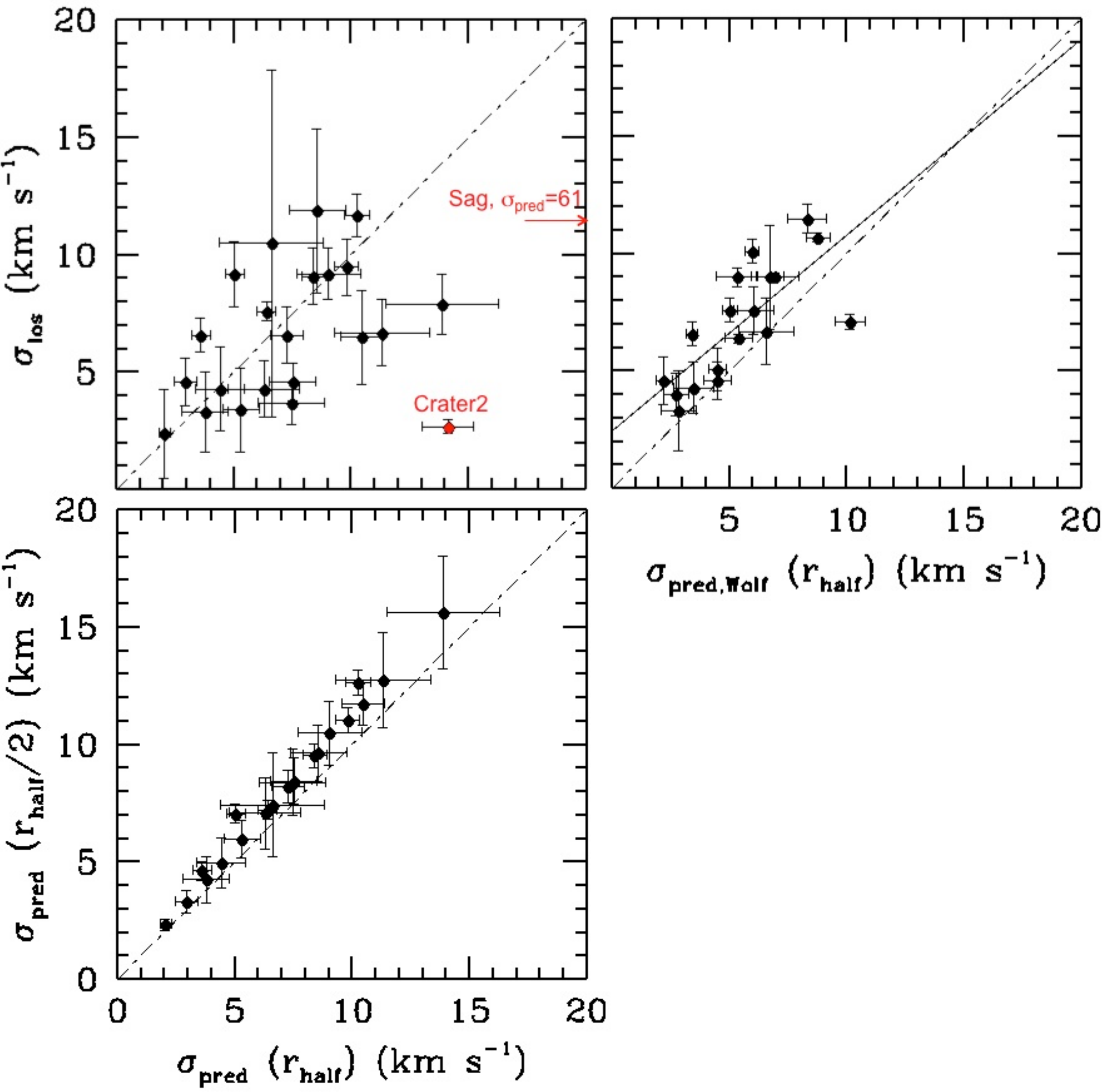}
\caption{Black points represent the dSphs of the VPOS. In each panel the short dash-long dash line represents the equality between predicted and observed $\sigma$. {\it(Top-left panel)}: observed $\sigma_{los}$ versus predicted value, $\sigma_{pred}$, which is the quadratic combination of the dispersion created by the Galactic force (Eq.~\ref{eq4}) with that caused by stellar mass. The latter is calculated through $\sigma_{dSph,stars}$ = $(M_{stellar}/(2\mu r_{half}))^{1/2}$, with $\mu$ = 580 $M_{\odot} pc^{-1} km^{-2} s^{2}$ \citep{Walker2009a}. The red point represents Crater 2, and the arrow indicates Sagittarius, for which $\sigma_{pred}$ are well offset (see value given in $km.s^{-1}$). {\it(Top-right panel)}: same as the left panel but for which $\sigma_{los}$ and $\sigma_{pred}$ values are coming or calculated from \citet{Wolf2010}. $\sigma_{pred}$ accounts for the full gravitational potential variations including those due to MW mass changes within the dSph volume and for the effect of the stellar mass (see Appendix~\ref{improved_calc}). Both quantities correlate with ñ$\rho$= 0.81, t= 5.5 and P= 2 $10^{-5}$, i.e., far better than do predictions shown from Eq.~\ref{eq4}. The solid line represents the best fit. {\it(Left-bottom panel)}: comparison of predicted values of $\sigma$ calculated (see Appendix~\ref{calc_rhalf/2}) at $r_{half}$/2 with that at $r_{half}$ from Eq.~\ref{eq4}.
}
\label{Fig3}
\end{figure*}

A strong argument in favor of DM-induced velocity dispersions comes from the flatness with radius of the observed velocity-dispersion profiles \citep{Strigari2008,Walker2009a,Wolf2010}. It could be argued as well that Eq. 5 is only predictive of values measured at $r_{half}$. In Appendix~\ref{calc_rhalf/2} we calculate the expected $\sigma_{los,MW}$ values at $r_{half}$/2, and bottom-left panel of Figure~\ref{Fig3} evidences that they are very similar to those at $r_{half}$. Numerical simulations of DM free galaxies by \citet{Yang2014} show that most of them may have flat velocity profiles after their first infall into the MW hot gas and gravitational potential (see their Fig. 6). Their elongated morphologies along their trajectories can also be predicted (see their Fig. 7 and 10 and also Appendix~\ref{simus}), and in Appendix~\ref{simus} (see Table~\ref{Tab3}) we show that they share similar velocity gradient properties than the observed ones.

The above calculations can be refined. For example it has been argued \citep{Wolf2010} that, to compare with half the stellar mass, the total mass is better estimated within a sphere with a radius of 4$r_{half}$/3, raising $\mu$ to 930 $M_{\odot} pc^{-1} km^{-2} s^{2}$. We have also neglected the possible variations of the MW mass within the dSph volume, and the potential variation $\Delta$$\phi$ estimated using Eq.~\ref{eq2} would be more accurate if replaced by $\Delta$$\phi$= $\phi$ ($\Delta$$M_{MW}$/$M_{MW}$ - $\Delta$$D_{MW}$/$D_{MW}$). Applying these changes (see Appendix~\ref{improved_calc}) modifies Eqs. 2 and 4 by multiplying their right sides by $\alpha$, which is a parameter depending on the shape of the MW mass profile (see Eq.~\ref{EqS18} in Appendix~\ref{improved_calc}). In Eq.~\ref{eq5}, the factor 3.529 is replaced by 5.657 $\alpha$ and the relation between $M_{tot}$ and its prediction ($M_{tot,pred}$) from $g_{MW}$ shows a scatter of only 0.18 dex over three decades, i.e., comparable to or even tighter than the fundamental (Tully-Fisher) relation between mass and velocity of spiral galaxies. Comparison between the top-right and top-left panels of Figure~\ref{Fig3} shows that the prediction is then improved, i.e., knowing $g_{MW}$ and $r_{half}$, one may predict quite accurately the observed line-of-sight velocity dispersions. Note also that the relation in the top-right panel of Figure~\ref{Fig3} has a slope almost equal to 1 after removing Sextans (the point with the largest predicted $\sigma$), the latter galaxy showing some discrepancy with other classical dSphs (see Appendix C). \\

In principle, the overall effect of Galactic acceleration should be integrated over the dSph orbital motions through the MW halo. However, during a first passage, one can consider that the Galactic acceleration instantaneously affects dSph kinematics since the crossing time ($t_{cross}$ $\sim$ $r_{half}$/$\sigma_{los}$) is ten to several hundreds times smaller than the encounter time ($t_{enc}$ $\sim$ $D_{MW}$/$V_{gsr}$) for dSphs lying within the VPOS (see Table~\ref{Tab1}). Then the overall dSph internal structures and kinematics are affected, and driven significantly out of equilibrium or, alternatively and for specific orbital parameters, could lead to quasi-stable satellites (see, e.g., \citealt{Casas2012}). That dSph galaxies are out of equilibrium has already been proposed \citep{Kuhn1989,Kuhn1993}, but objected to by \citet{Mateo1993} who questioned how we could observe such systems together, since they should disperse within short timescales. The VPOS along which most dSph PAs are aligned suggests an ordered spatial distribution and motions of the dSphs together with the Magellanic Clouds during their first approach, which explains the close relationship between $g_{MW}$ and $\sigma_{los}^2$ shown in Figures 2 and 3.\\

\begin{figure}
\epsscale{0.5}
\plotone{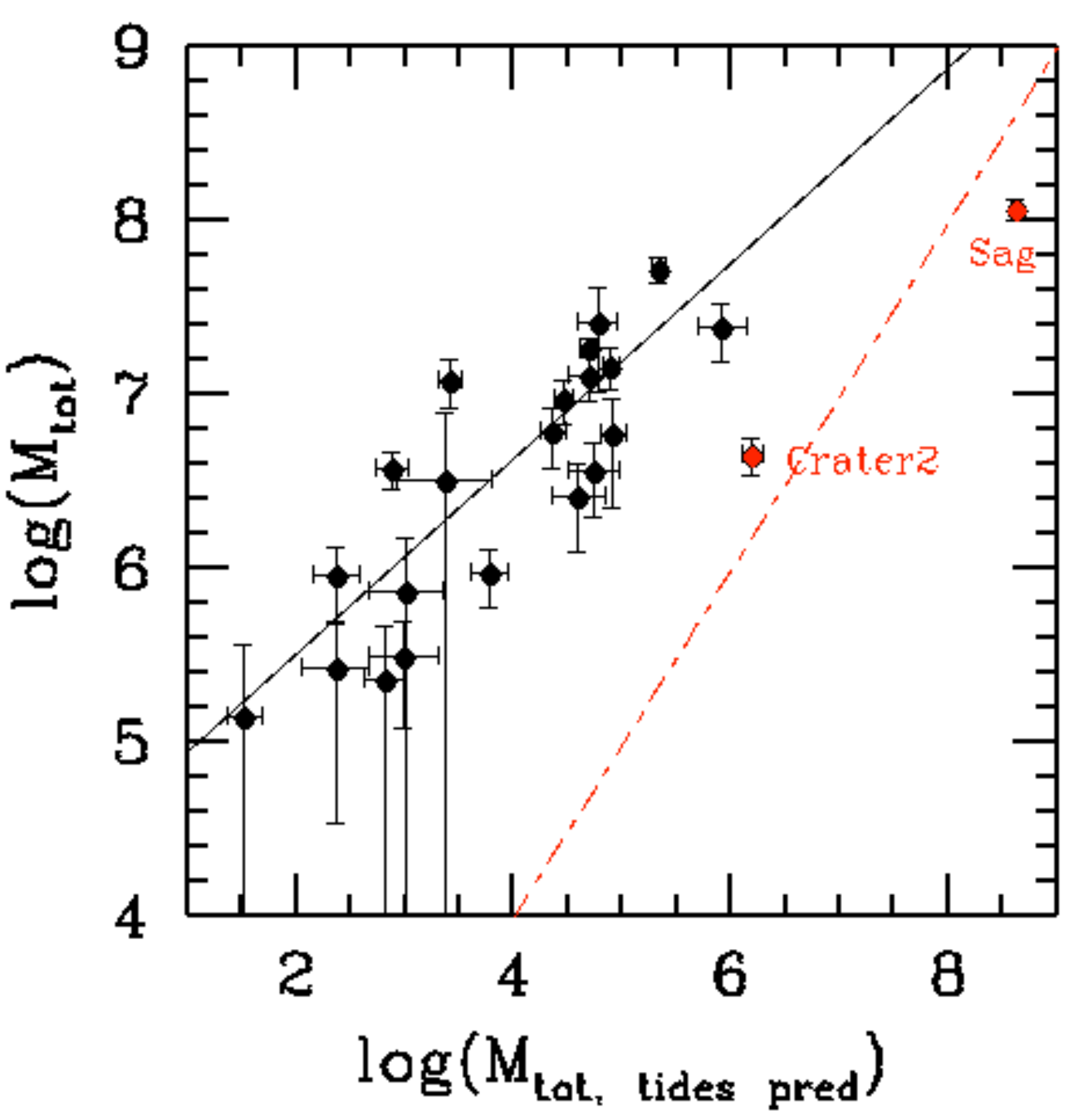}
\caption{Total mass versus that predicted if Galactic tidal forces are ruling the star orbital motions in a satellite (see Appendix~\ref{calc_tidal}). Black points represent the dSphs of the VPOS, red points represent Crater 2 and Sagittarius, respectively. The solid line shows the best fit of the black points with $\rho$= 0.83, t= 6.6 and P= 7.7 $10^{-7}$, i.e., strong but slightly less impressive correlation than that in the bottom-left of Figure~\ref{Fig2}. The red full line represents equality and passes very near Sagittarius.
}
\label{Fig4}
\end{figure}

Is this proof that the VPOS dSphs are on their first passage? The answer is probably given by Sagittarius, which is well offset from the relations drawn by the VPOS dSphs in Figures 2 and 3. It has experienced at least two and perhaps up to five passages \citep{Dierickx2017} at pericenter. Let us then consider a system fully dominated by the Galactic tidal forces and calculate the predicted total mass (see Appendix~\ref{calc_tidal}) as we did for the Galactic acceleration. This leads to Figure~\ref{Fig4} and it evidences that the MW tidal forces cannot account for the total mass (or $r_{half}$ $\sigma_{los}^2$) of dSphs by two to four decades, while it matches that of Sagittarius. For the latter, the encounter time is almost equal to the characteristic crossing time ($t_{enc}$/$t_{cross}$ = 0.82): it suggests that the MW tidally locked Sagittarius as the Earth did to the Moon.

\section{Conclusion: What can be derived from the kinematic studies of the MW dSphs?} \label{sec:discussion_conclusion}

Eqs. (2)-(5) have been established without any assumption on the mass of dSphs, and Figure~\ref{Fig2} shows that the MW acceleration predicts well the amplitude of the observed $r_{half}$ $\sigma_{los}^2$ (or $M_{tot}$). It does not mean that there is no DM in the MW dSphs, though its amount cannot be deduced from $r_{half}$ $\sigma_{los}^2$ values. There is no more argument for the very high DM fractions ($>>$10) in the faintest dSphs, thus questioning the search for DM in these evanescent objects. Furthermore, it is no longer justified that MW dSphs follow the universally adopted relationship between the stellar and total masses in the dwarf regime, which is therefore put into question. A similar effect is expected for the radial acceleration relation (see, e.g., \citealt{Lelli2017}) for which the dwarf regime has been populated mostly by both MW and M31 dSphs.\\

We verified that a MW mass model based on its rotation curve \citep{Sofue2012} increases the correlation strengths found in Figure~\ref{Fig2}. Since the dSphs lie at distances ranging from 20 to 250 kpc, the actual mass profile of the MW can be probed at large distances for which rotation curve measurements may lead to some ambiguous results. \\

Figures~\ref{Fig2} to~\ref{Fig4} show that the enigmatic ultra-faint dwarf Crater 2 \citep{Caldwell2017} shares many properties with Sagittarius, implying also an early infall. This opens a new avenue for dating the epoch of infall for the individual dSphs. A first passage for the Magellanic Clouds and most dSphs is consistent with the formation of the HI Magellanic System \citep{Hammer2015}, the first two possibly being responsible of the Magellanic Stream and its double filamentary structure, and the others creating the four Leading Arm structures. Finally, the dispersion resulting from the motions perpendicular to the dSph motion on the sky should be significantly smaller than $\sigma_{los}$ (dominated by MW force) and than the dispersion along the dSph trajectory (expansion through time-integrated MW tidal action). From our simulations (see Appendix~\ref{simus} and the online animated Figure~\ref{fig5anim}) we estimate that, in the absence of initial rotation, the former component can be $\sim$ 1/5-1/3 times the others, resulting into a pancake shape for these objects (see the online animated Figure~\ref{fig5anim}), a prediction to be verified with the GAIA future data releases \citep{Gaia2016}.\\

\begin{figure}
\epsscale{0.9}
\plotone{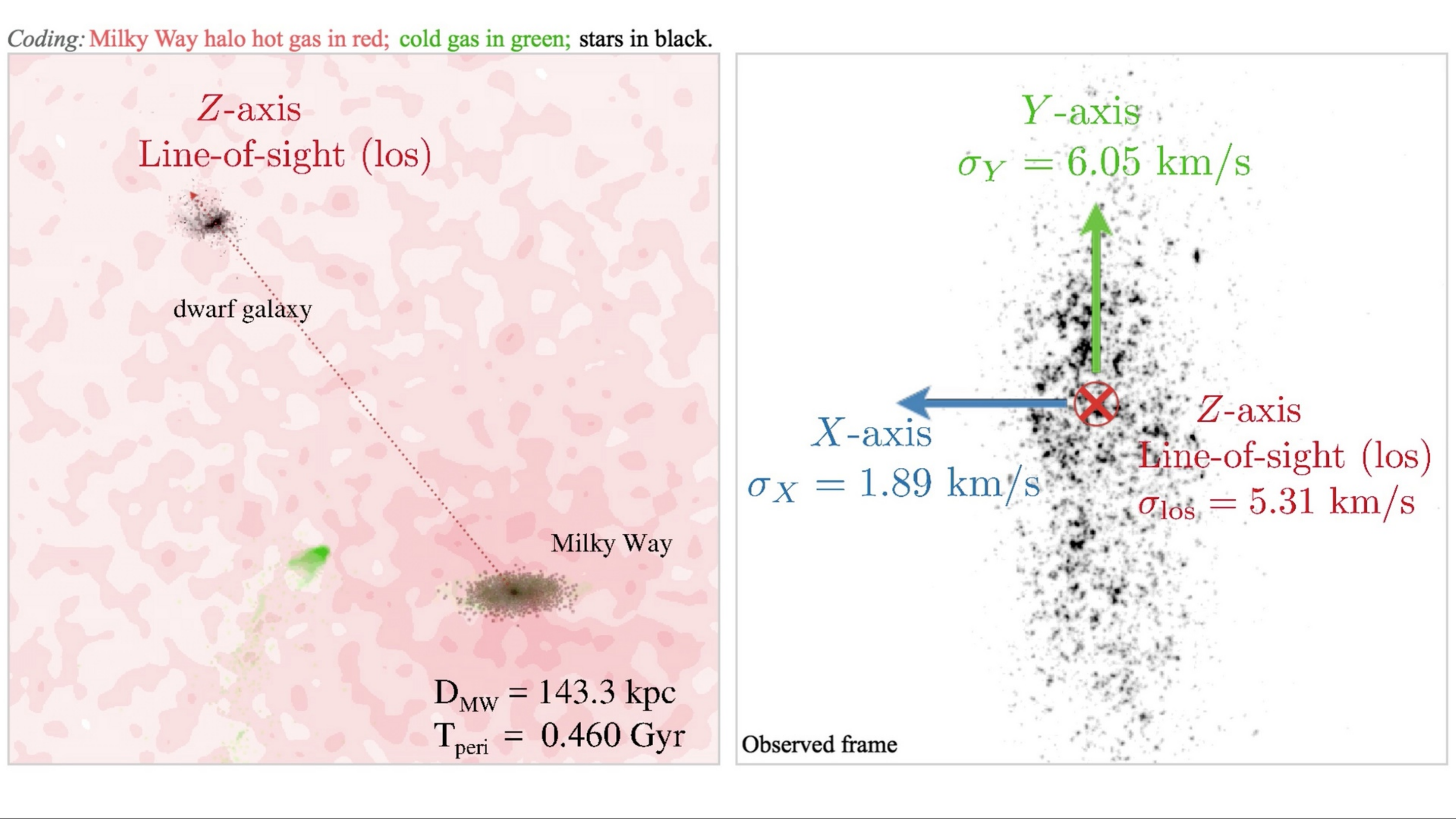}
\caption{The online animated Figure shows the transformation of a gas-rich dwarf (see the second line of Table~\ref{Tab3}) into a dSph after a first passage into the MW halo. The left panel shows the orbital plane with both the initially gas-rich galaxy (HI gas in green, stars in black) and the MW and its halo hot gas (in red).  The right panel shows how an observer located at the Earth would observe the dwarf galaxy. The above figure illustrates the 3D $\sigma$ properties at $T_{peri}$= +0.46 Gyr.   At the end the dSph is rotated for a better view of its 3D shape. } \label{fig5anim}
\end{figure}

In principle, studies of dwarfs surrounding external galaxies could be an important test of the dwarf DM content. M31 dwarfs might be affected similarly to MW dwarfs. This is because dwarfs surrounding the M31 halo either belong to the M31 gigantic disk of satellites  \citep{Ibata2013} or are expected to lie around the M31 disk. Since both structures are seen edge-on, it increases the contribution to either the sigma component caused by the M31 acceleration or that created by the accumulation of tidal effects. To circumvent these projection effects, one would have to look for another galactic halo gigantic structure, which would not be seen edge-on. We consider whether the very recent discovery of NGC1052-DF2 without any sign of DM (see, e.g., \citealt{van Dokkum2018}) could be an illustration of these effects. Studying more similar examples as well as fully isolated galaxies would be an interesting follow-up to verify or disprove the existence of two populations of dwarfs (see, e.g., \citealt{Kroupa2012}).\\

{\it Note Added in Proof}: Just after the present paper was accepted for publication, several papers (Gaia collaboration, A. Helmi, et al. ArXiv:1804.09381 and Fritz et al., ArXiv :1805.00908) appeared and discussed the exact structure of the VPOS. We note that the results obtained in the present paper apply to all MW dSphs whether they lie or not in the VPOS (see in particular Sagittarius, Hercules or Bootes, which lie or may lie outside the VPOS).\\

\acknowledgments
We are very grateful to Piercarlo Bonifacio, Pavel Kroupa, Marcel Pawlowski, Gerhard Hensler, Gary Mamon and Beatriz Barbuy who read preliminary versions and provided us with useful comments and references. We are indebted to Matthew Walker who kindly provided us with additional data to allow us with a full analysis of the enigmatic Ursa Minor dwarf spheroidal. We thank the referee for his/her useful comments and suggestions. The China-France International Associated Laboratory ÒOriginsÓ has supported this work. J. L. W. thanks the China Scholarship Council (NO.201604910336) for the financial support. 
\newpage

\appendix
\section{ Estimation of the velocity gradients of 8 dSph galaxies.}
\label{vel_grad}
We are trying to determine whether the stellar radial velocities in the dSphs tend to be oriented along any preferential direction. For this purpose we computed the correlation between radial velocities and sky coordinates, using the robust Kendall's $\tau$ nonparametric measure of the degree of correlation \citep{Kendall1938}. Stars were selected using membership criteria provided by references given in Table~\ref{Tab2} (last column). This tablealso lists the angle $\theta$ from the Galactic plane, which leads to the most significant correlation between the position (X) of the stars after a $\theta$ rotation and their radial velocity. The significance is provided by the p-value of the test, the null hypothesis being that velocities and positions are uncorrelated.$\theta$ is provided only when the correlation test is significant, with a probability of less than 1\% to be uncorrelated. The uncertainty on this angle has then been estimated using bootstrap resampling.\\

\begin{figure}
\epsscale{0.9}
\plotone{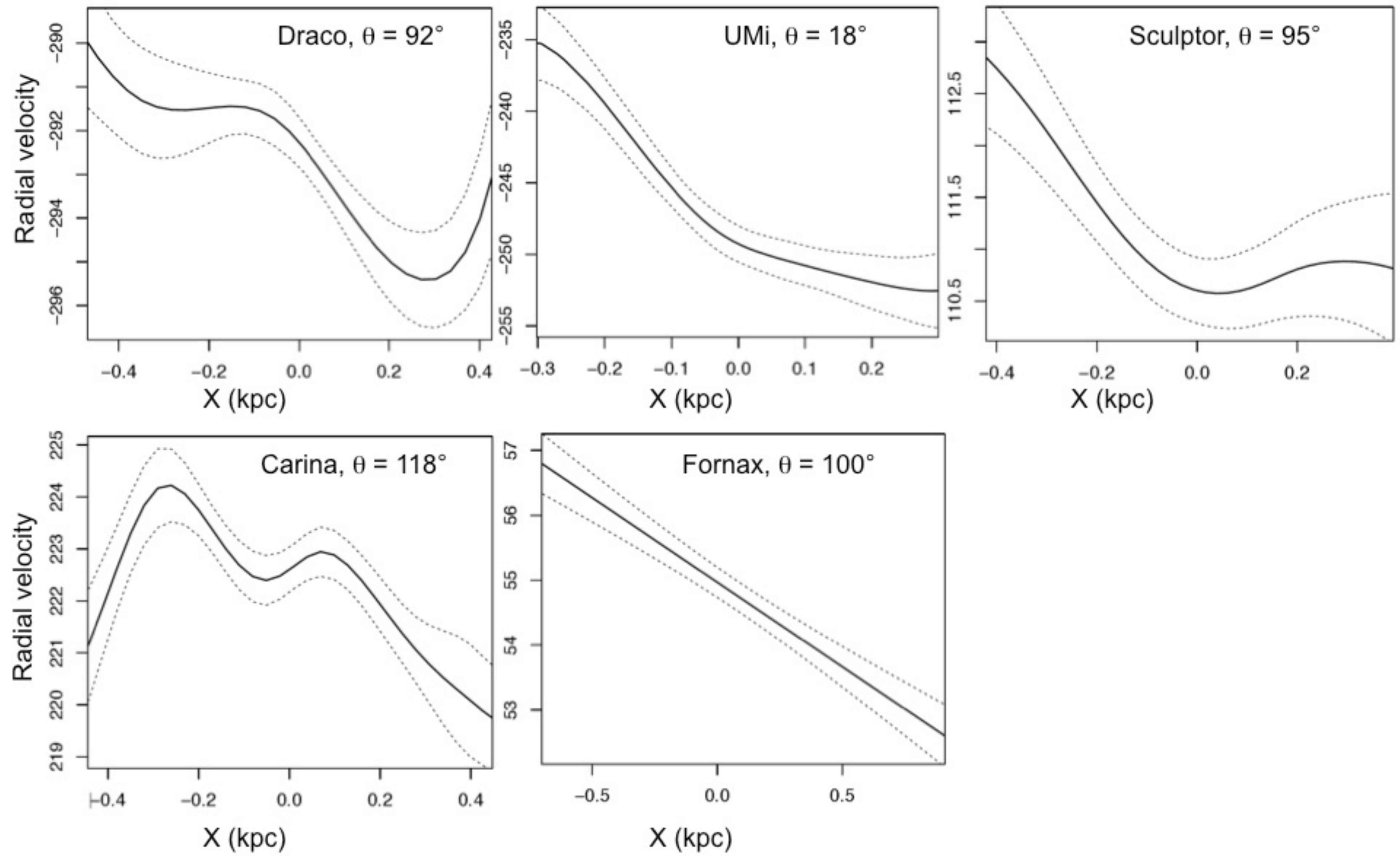}
\caption{Radial velocities vs position after a rotation $\theta$ indicated in Table~\ref{Tab2} for the five dSphs (Draco, UMi, Sculptor, Carina, Fornax) for which the velocity gradient is significant. In order to provide a trade-off between the goodness-of-fit, i.e. roughness, of the noisy data and the required smoothness, the data are represented by cubic smoothing splines \citep{Helwig2015} with 12 knots, using a robust 3$\sigma$ clipping implementation. The left and right 5\% of the abscissa data are not shown. In dashed lines, the $\pm 1\sigma$ curves represent the posterior standard uncertainties of the fitted values.
}
\label{FigS1}
\end{figure}

Among the eight most massive dSphs belonging to the VPOS, for five of them we have been able to find a significant velocity gradient. These include Carina, Fornax, and Sculptor for which our results and data are essentially the same as those of \citet{Walker2008}. The number of stars shown in Table~\ref{Tab2} is sufficiently high to robustly sample the velocity measurements. Figure~\ref{FigS1} shows the relation between the radial velocity and the position along the galaxy, whose orientation is given by $\theta$. All but UMi show a velocity gradient oriented along the VPOS, i.e., with $\theta$$\sim$ 90\degr. This further suggests tidal effects linked to the MW gravitational potential. \

\begin{deluxetable*}{ccccccc}
\tablecaption{Velocity gradients for classical dSphs\label{Tab2}}
\tablewidth{0pt}
\tablehead{
\colhead{dSph} & \colhead{N(stars)} & \colhead{$\theta$} & \colhead{P-value} & \colhead{2$\times r_{half}$} & \colhead{$\Delta$X} & \colhead{References for star}\\
\colhead{} & \colhead{} & \colhead{(degrees)} & \colhead{} & \colhead{(kpc)} & \colhead{(kpc)} & \colhead{membership}
}
\decimalcolnumbers
\startdata
Draco &  581 &  92 $\pm$27 &  0.003 &  0.4 &  0.84 &  \citet{Walker2007,Kleyna2002,Walker2015}\\
UrsaMinor &  425 &  18 $\pm$25 &  0.005 &  0.56 &  0.6 &  \citet{Armandroff1995}, Walker, private communication\\
Sculptor &  1369 &  95 $\pm$6 &  0.008 &  0.52 &  0.8 &  \citet{Walker2008}\\
Sextans &  533 &  - &  0.11 &  1.36 &  - &  \citet{Walker2008,Battaglia2011}\\
Carina &  939 &  118 $\pm$34 &  0.002 &  0.48 &  0.9 &  \citet{Walker2008,Munoz2006}(28, 37)\\
Fornax &  2516 &  100 $\pm$3 &  2$\times10^{-7}$ &  1.34 &  2.2 &  \citet{Walker2008}\\
LeoII &  239 &  - &  0.09 &  0.3 &  - &  \citet{Spencer2017}\\
LeoI &  400 &  - &  0.24 &  0.49 &  - &  \citet{Mateo2008,Sohn2017}
\enddata
\tablecomments{This table gives for each dSph (1st column) the number of stars (2nd column) used to test the Galactic angle $\theta$ of the dSph radial velocity orientation (3rd column). The latter corresponds to the angle providing the smallest p-value in the Kendall correlation test between the rotated position and the radial velocity, the corresponding p-value being provided in the 4th column. 5th and 6th columns provide the half-light diameter and the extent of the velocity gradient, respectively. The last column gives references from which stars have been pre-selected.}
\end{deluxetable*}

Alternatively it has been proposed \citep{Walker2008} that dark-matter dominated galaxies can be considered as solid bodies, and that their proper motions may create an apparent velocity gradient, due to the different perspective viewpoints toward an extended object. Such a "perspective rotation" could help to indirectly estimate proper motions (PMs). This results in a remarkable agreement for Fornax PMs, and indeed its velocity profile follows a straight line (see Figure~\ref{FigS1}) and may suggest a solid body rotation (however, see \citealt{del Pino2017} who support  a far more complex kinematics for this galaxy). Besides this, it is acknowledged \citep{Walker2008} that by integrating stars further beyond the half-light-radius, their PM determination for, e.g., Carina, becomes inconsistent because of tidal streaming motions. Table~\ref{Tab2} indicates that for Draco, UMi, Sculptor and Carina, our velocity gradient is extracted over a region ($\Delta$X) with a diameter that is 3-4 times the half light radius, supporting the association of the velocity gradient to tidal streaming motions, though it cannot be excluded that both effects are at work. The behavior of their velocity profiles is also not well represented by a solid body.

\section{Calculations of the MW potential gradient}
\label{calc}
\subsection{Calculations of the MW potential gradient at a projected radius R=$r_{half}$}
\label{calc_rhalf}
\begin{figure}
\epsscale{0.9}
\plotone{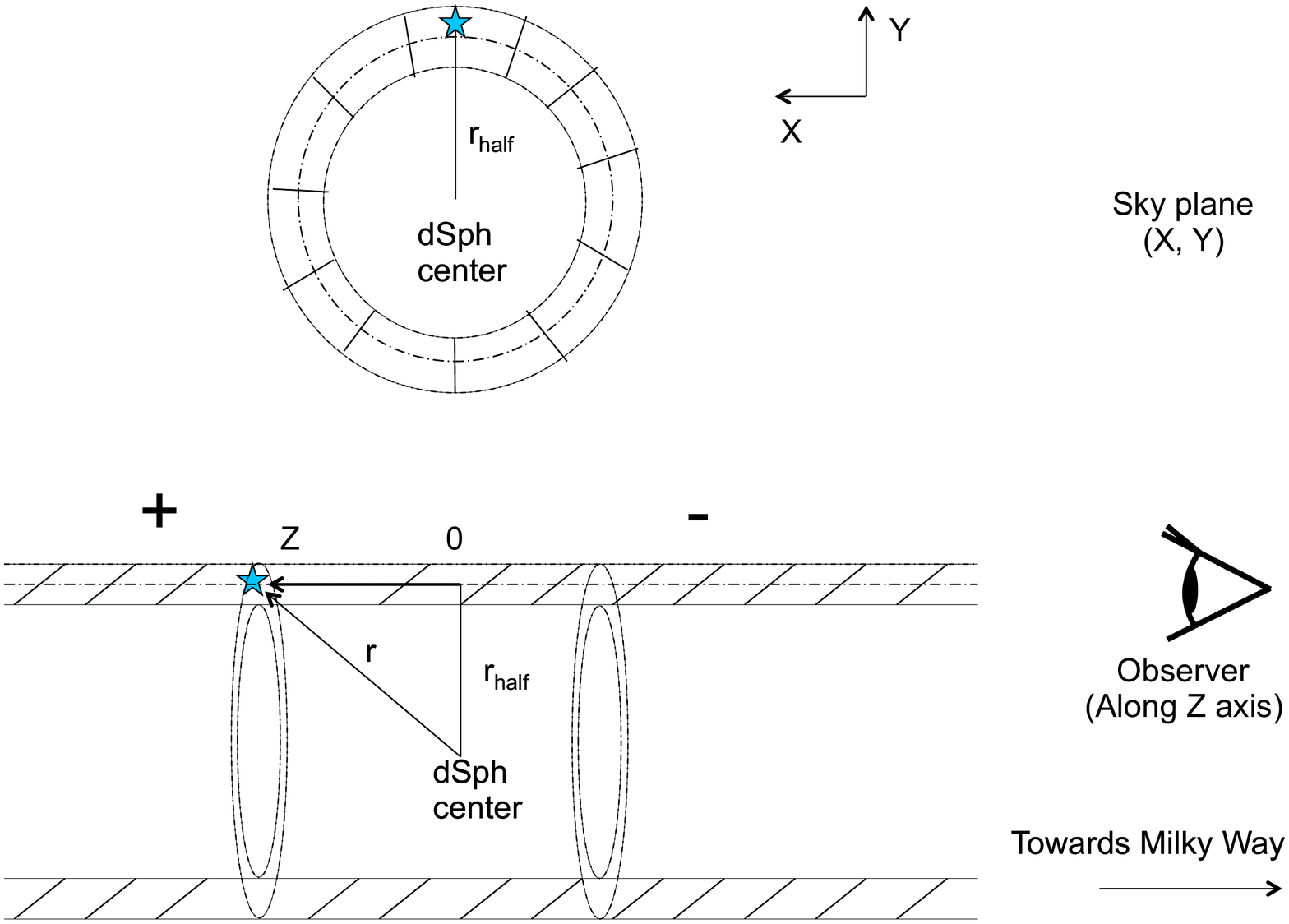}
\caption{Sketch illustrating the selection of stars in a tube at projected R= $r_{half}$ within a dSph galaxy, based on \citet{Walker2009a}. (Top): stars observed in the sky plane (X, Y) within a circular annulus centered on the projected radius R=$r_{half}$ (hatched area). (Bottom): projection along the line of sight (Z-axis) showing the position of the observer at the MW and the hatched areas that include observed stars. In fact, selected stars by \citet{Walker2009a} are confined to a tube that can be represented as concentric ellipses that are side views of the circular annulus shown in the top panel. This explains that at projected radius R=$r_{half}$, one may integrate velocity variations along the Z-axis, which may be related to MW potential variations between both half sides. Sign + and -, indicate the two regions where potential $\phi^+$ and $\phi^-$ have been averaged.
}
\label{FigS2}
\end{figure}

Let us consider that the dSph stars are distributed in a Plummer sphere, for which the density is given as follows:

\begin{equation}
   \rho  \left( r \right) = \frac{3M_{stellar}}{4  \pi  r_{half}^{3}}  \times   \left( 1+ r^{2}/r_{half}^{2} \right) ^{-\frac{5}{2}} \: \label{EqS1}
\end{equation}

where $M_{stellar}$ is the stellar mass of the dSph. Let us considere further the observations at a projected radius R=$r_{half}$ for which stars are selected on the sky within a circular annulus \citep{Walker2009a}. When de-projected along the line of sight, stars are actually confined to a tube, as shown by the bottom panel of Figure~\ref{FigS2}. One may consider the half-tube containing the stars that are farthest from the MW and calculate the average potential exerted by the MW:

\begin{equation}
< \phi^{+}> = \frac{-GM_{MW}}{D_{MW}} \frac{\int_{0}^{+\infty} \rho  \left( r \right)  \left( 1-Z/D_{MW} \right)  dZ }{  \int _{0}^{+\infty} \rho  \left( r \right)  dZ } \label{EqS2})
\end{equation}

This assumes Z/$D_{MW}$ $<<$ 1 and then 1/($D_{MW}$+Z) $\approx$ (1 - Z/$D_{MW}$)/$D_{MW}$. Since $r^2= r_{half}^2 + Z^2$, this results in:

\begin{equation}
  < \phi^{+}> =k \int_{0}^{+\infty} \left( 1-Z/D_{MW} \right)  \left( 1+ Z^{2}/ \left( 2r_{half}^{2} \right)  \right) ^{\frac{-5}{2}} dZ ,
  \label{EqS3}
\end{equation}
with
$ k= - \left( \frac{GM_{MW}}{D_{MW}} \times  \frac{3M_{stellar}}{16\times \sqrt[]{2}  \pi  r_{half}^{3}} /  \int _{0}^{+\infty} \rho  \left( r \right)  dZ \right)  $

We do not calculate k, which is assumed to be a constant since we ignore for the moment the variation of the MW mass ($M_{MW}$) within the dSph volume. One may also calculate $< \phi^{-} >$, i.e., the average gravitational potential exerted by the MW on the dSph stars within the closest half-tube, by using the same integral, but now integrating from Z=-$\infty$ to 0. This gives:

\begin{equation}
 < \phi ^{-}> =k \int _{-\infty}^{0} \left( 1-Z/D_{MW} \right)  \left( 1+ Z^{2}/ \left( 2r_{half}^{2} \right)  \right) ^{\frac{-5}{2}} dZ  
 \label{EqS4}
\end{equation}

The constant k is the same as in Eq.~\ref{EqS3} because integration of the density from Z=-$\infty$ to 0 is equal to that from Z= 0 to +$\infty$. We find:\\

$ < \phi^{+}> =2k r_{half}^{2}  \left( \sqrt[]{2}\frac{D_{MW}}{r_{half}}-1 \right) $ and  
$ < \phi^{-}> =2k r_{half}^{2}  \left( \sqrt[]{2}\frac{D_{MW}}{r_{half}}+1 \right)   $\\

The MW gravitational potential relative variation is given by:

%\begin{equation}
%  \frac{ \Delta  \varnothing }{ \varnothing }=\frac{< \phi^{-}> - < \phi^{+}>}{< \phi^{+}> + < \phi^{-}>}= \frac{r_{half}}{\sqrt[]{2}D_{MW}}   (Eq. S5)
%\end{equation}
\begin{equation}
  \frac{ \Delta  \phi }{ \phi }=\frac{< \phi^{-}> - < \phi^{+}>}{< \phi^{+}> + < \phi^{-}>}= \frac{r_{half}}{\sqrt[]{2}D_{MW}}   
  \label{EqS5}
\end{equation}

Then Eq.~\ref{EqS5} results in Eq.~\ref{eq2}.  In principle, integration of Z to infinity would violate the assumption that Z/$D_{MW}$ $<<$ 1, but in practice the profile is sufficiently steep so that integrating Z to a small fraction of $D_{MW}$ instead of infinity does not affect the result. We have also verified that adopting other (steep) profiles provides quite similar results. For example, adopting a perfect sphere with $\rho$ (r) $\sim$ $(1 + r^2/b^2)^{-2}$, one finds $b^2= r_{half}^2/3$, and then $\frac{ \Delta  \phi }{ \phi }= \frac{4 r_{half}}{ \pi \sqrt[]{3 }D_{MW}}=1.0396 \frac{r_{half}}{\sqrt[]{2 } D_{MW}}  $. Changing the density profile therefore impacts the results only by a numerical factor very close to 1. Since a Plummer sphere was adopted by \citet{Walker2009a} in their analyses, we adopt Eq.~\ref{EqS5} and Eq.~\ref{eq2} in the following.

\subsection{Calculations of the expected $\sigma_{los,MW}$ values at $r_{half}$/2}
\label{calc_rhalf/2}

For stars observed at a projected radius R= $r_{half}$/2, one may calculate the potential variation by adopting $r^2= (r_{half}/2)^2 + Z^2$ in Eqs. ~\ref{EqS1} to ~\ref{EqS5}. This leads to:

\begin{equation}
\frac{ \Delta  \phi }{ \phi }=  \frac{< \phi ^{-}> - < \phi ^{+}>}{< \phi ^{+}> + < \phi ^{-}>} = \frac{\sqrt[]{5} r_{half}}{4 D_{MW}} 
\label{EqS11}
\end{equation}

This would induce velocity dispersions given as follows:\\

$ \sigma_{los,MW}^{2} \left( \frac{r_{half}}{2} \right) =\frac{4 G M_{MW} \left( D_{MW} \right)   \times  r_{half}}{ \sqrt[]{5} D_{MW}^{2}} $\\

We also calculate the velocity dispersion caused by the stellar mass content of the dSph, $\sigma_{dSph,stars}$, following Eq.~10 of \citet{Walker2009a}:

\begin{equation}
\sigma_{(dSph,stars)}^2=5 M_{stellar} (\frac{r_{half}}{2})/(\mu r_{half})				
\label{EqS12}
\end{equation}

where $\mu$ = 580 $M_{\odot} pc^{-1} km^{-2} s^{2}$. The stellar mass within $r_{half}$/2 can be deduced from Figure~\ref{Fig2} of \citet{Walker2007} for which the mass profile can be estimated between 10 pc to $r_{half}$:

\begin{equation}
log(M_{stellar}(r) ) \approx ˜0.854 ×log(r)+ 2.593			
\label{EqS13}
\end{equation}

\begin{equation}
M_{stellar}(\frac{r_{half}}{2})˜\approx 0.553 × M_{stellar} (r_{half} )				
\label{EqS14}
\end{equation}

A quadratic combination provides the predicted $\sigma$ at $r_{half}$/2, which is the ordinate of the bottom-left panel of Figure~\ref{Fig3}:

\begin{equation}
\sigma_{pred} (\frac{r_{half}}{2})= \sqrt{\sigma_{(los,MW)}^2+\sigma_{(dSph,stars)}^2 }  				
\label{EqS15}
\end{equation}

At $r_{half}$/2, the correlation between $\sigma_{los}$ and $\sigma_{pred}$ is slightly more significant ($\rho$= 0.61, t= 3.4 and P= 1.4 $10^{-3}$) than at $r_{half}$. A constant value for $\sigma_{los}$ expected from Figure~\ref{Fig3} is also in agreement with expectations from simulations \citep{Yang2014} and with the suggestion that at all radii, stellar kinematics is intimately affected by the Galactic acceleration.

\subsection{Improved calculations for total mass and MW potential variations}
\label{improved_calc}
According to \citet{Wolf2010} the total mass is optimally calculated within $r_{1/2}Ê\approx (4/3) r_{half}$, where $r_{1/2}$ and $r_{half}$ are the 3D de-projected and the 2D projected half-light radii, respectively. This leads to $M_{tot}(4r_{half}/3)= \mu r_{half} \sigma_{los}^2$ where $\mu \approx 930 M_{\odot} pc^{-1} km^{-2} s^{2}$ and $\sigma_{los}$ is the observed line-of-sight velocity dispersion. We followed \citet{Wolf2010} by adopting values from their Table 1 for 
calculating the total-to-stellar mass ratios as $2\times M_{tot}(4r_{half} /3)/M_{stellar}$. Notice that this limits the sample to 18 galaxies instead of 21 from \citet{Walker2009a}. Furthermore, let us replace Eq.~\ref{eq2} by $\Delta \phi= \phi (\Delta M_{MW}/M_{MW} - \Delta D_{MW}/D_{MW})$, which accounts for the variation of the MW mass within the dSph volume. We need then to use the \citet{Sofue2012} formulae (see their Eq.~17) assuming $\Delta D_{MW}$= Z (see Figure~\ref{FigS2}):

\begin{equation}
\frac{\Delta  M_{MW}}{Z}= 2.6 \times 10^{11}  \frac{D_{MW}}{ (12.5 +D_{MW})^2}  
\label{EqS16}
\end{equation}

In Eq.~\ref{EqS16} distances and masses are in kpc and solar mass units, respectively. The MW potential can be developed as follows:

\begin{equation}
\phi(Z) \approx  - \frac{G˜ M_{MW}}{D_{MW}}  \times  (1+ \frac{\Delta M_{MW}}{M_{MW}}) (1-\frac{Z}{D_{MW}} )					
\label{EqS17}
\end{equation}

It results that the variation of the potential between 0 and Z is:

\begin{equation}
\Delta  \phi(Z) \approx \phi( 1 - \alpha Z/ D_{MW}) 
\label{EqS18}
\end{equation}

where 
$ \alpha = 1-((2.6 \times 10^{11})/M_{MW} )× (D_{MW}/(12.5+D_{MW}) )^2$

Notice that $\alpha$ ranges between 0 and 1, with values from 0.29 (Segue) to 0.605 (Leo I), and that it can be calculated again for any type of MW mass profile. To calculate the average value of $\Delta \phi$ one needs to integrate Eq.~\ref{EqS18} along the X-axis at the projected radius $r_{half}$ (see Figure~\ref{FigS2}); assuming a Plummer profile (Eq.~\ref{EqS1}), one may calculate the average potential for the closest and farthest half of the stars (relative to the MW) as in Eqs. \ref{EqS3} and \ref{EqS4}, respectively, and then:

\begin{equation}
< \phi ^{+}> =k \int _{0}^{+\infty} ( 1- \alpha Z/D_{MW} ) ( 1+ Z^{2}/( 2r_{half}^{2} ) )^{\frac{-5}{2}} dZ  
\label{EqS19}
\end{equation}

In Eq.~\ref{EqS19}, k has the same definition as in Eq.~\ref{EqS3} and $< \phi^{-} >$ can be calculated by using the same integral, but from Z=-$\infty$ to 0. Following Eq.~\ref{EqS5}, we find:

\begin{equation}
\frac{ \Delta  \phi}{ \phi}=  \frac{< \phi ^{-}> - < \phi ^{+}>}{< \phi ^{+}> + < \phi ^{-}>} = \frac{\alpha \; r_{half}}{\sqrt{2} \; D_{MW}}
\label{EqS20}
\end{equation}

Assuming that the specific kinetic energy is induced by the potential variations (Eq.~\ref{eq3}), this leads to:

\begin{equation}
\sigma_{los,MW}^2=  \sqrt{2} \; \alpha\;  G\;   \frac{M_{MW}}{D_{MW}^2 } \;  r_{half}  = \sqrt{2} \: \alpha\;  g_{MW} \: r_{half} \: 
\label{EqS21}
\end{equation}

The dispersion due to the MW is then combined in a quadratic manner with that caused by the stellar mass to lead to the predicted value of $\sigma$ in the top-right panel of Figure~\ref{Fig3}. To represent better stellar populations of dSphs \citep{Lelli2017} we have used stellar mass with $M_{\odot}/L_{\odot}$=2 (instead of 1 as adopted throughout the text) in V-band, and it indeed further improves the correlation.  One may deduce the total mass predicted if the MW gravitational force dominates the velocity dispersion:

\begin{equation}
M_{tot,pred}(\frac{4 r_{half}}{3})˜\approx 5.657 \; \alpha \; M_{MW}(D_{MW})  \times (\frac{r_{half}}{D_{MW} })^2  			
\label{EqS22}
\end{equation}

\begin{figure}
\epsscale{0.9}
\plotone{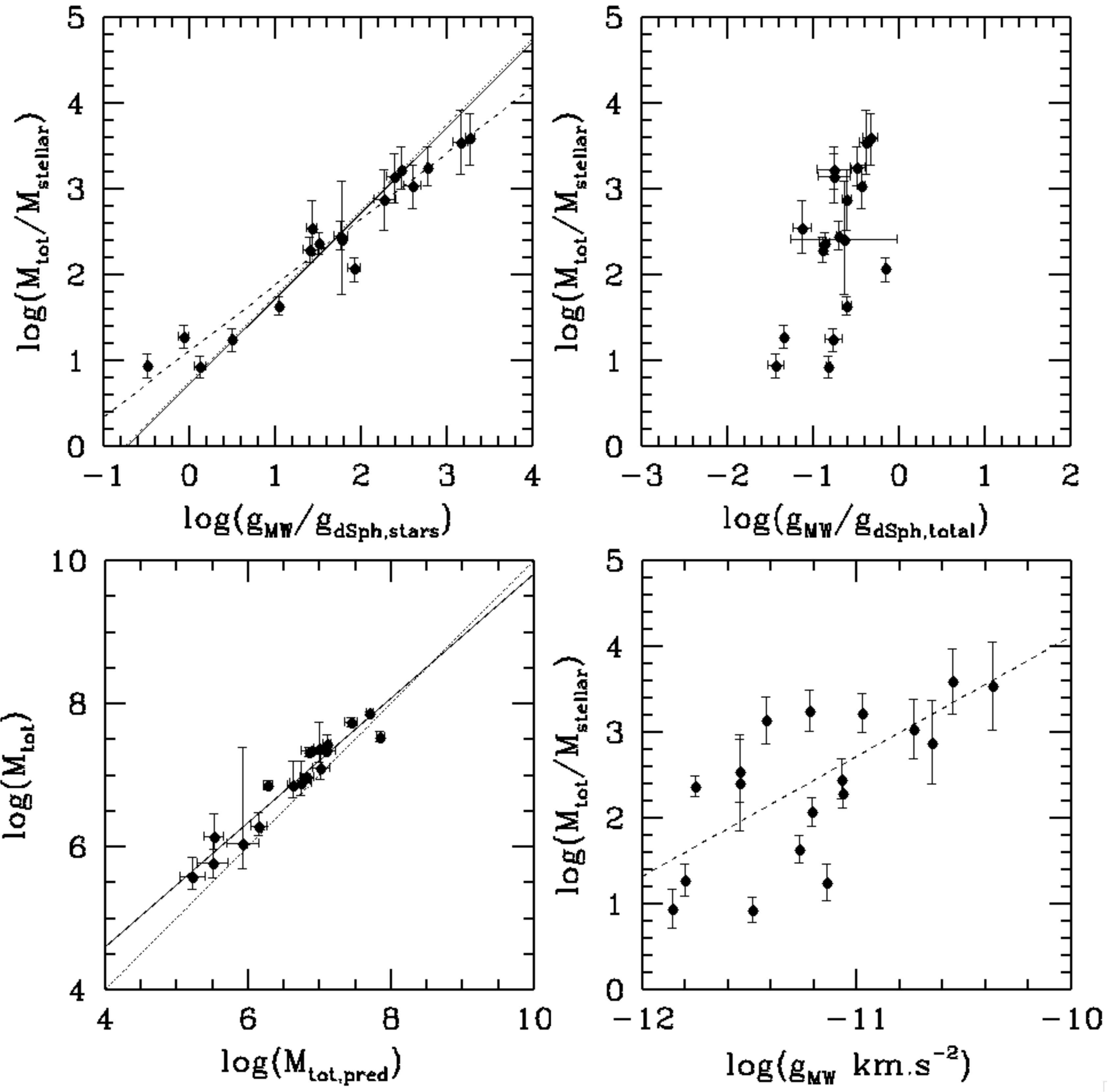}
\caption{Same panels as Figure~\ref{Fig2}, but replacing the estimate of $M_{tot}(r_{half})$ by that of $M_{tot}(4r_{half}/3)$ following \citet{Wolf2010}, and by calculating the predicted total mass, $M_{tot,pred}$, using Eq.~\ref{EqS22} instead of Eq.~\ref{eq5}. Correlation strengths for the 18 VPOS dSphs (black points) are from top-to-bottom, left-to-right: $\rho$= 0.94, t= 10.8 and P$< 10^{-10}$, $\rho$= 0.56, t= 2.7 and P= 0.007, $\rho$= 0.975, t= 17.6 and P$<< 10^{-10}$, and, $\rho$= 0.62, t= 3.2 and P= 3 $10^{-3}$. In the top-left panel one may hardly distinguish the best fit ($\chi^2$ minimization, full line) from expectations from Eq.~\ref{EqS22} (see the dotted line slightly above the full line).
}
\label{FigS6}
\end{figure}

Eq.~\ref{EqS22} and Figure~\ref{FigS6} replace Eq.~\ref{eq5} and Figure~\ref{Fig2}, respectively, after adopting the above changes. In the top-left panel of Figure~\ref{FigS6}, minimizing $\chi^2$ for a line with a slope of 1 provides a factor of 5.495 instead of 5.657 in Eq.~\ref{EqS22}. 

Let us examine the bottom-left panel of Figure~\ref{FigS6}, for which the ordinate represents the total (dynamical) mass, which includes the stellar contribution. This justifies the need to add to the abscissa (Eq.~\ref{EqS22}) the stellar mass under the same assumptions as above ($M_{\odot}/L_{\odot}$=2). This leads to a robust relationship with a very small scatter (0.18 dex) over three decades, which is even tighter than the best-established Tully-Fisher relation \citep{Reyes2011}. The dynamics of MW dwarfs are governed by the MW gravitational acceleration, confirming the robustness of the impulse approximation \citep{Binney1987}. We further note that the points are slightly above the equality between ordinate and abscissa. Since our calculations are valid whatever is the dSph total mass, one may investigate whether dark matter could cause this small discrepancy.\\

Alternatively one may try to optimize the strength of the correlation through Eq.~\ref{EqS22}, which depends on the MW mass profile through $\alpha$ and $M_{MW}(D_{MW})$. This potentially provides an independent method for calculating the mass profile and then the total mass of the MW. Kinematic data from \citet{Walker2009a} are invaluable measurements of $\sigma_{los}$ given the adopted methodology and since velocity gradients have been subtracted from raw data. One needs to investigate further, when they are combined with $r_{half}$ estimates, whether they are sufficiently accurate to perform the tests described above.

\section{Calculations of, and correlations with the Milky Way tidal acceleration}
\label{calc_tidal}
Let us consider the tidal acceleration, $ g_{MW,tides}  \approx G M_{MW}(D_{MW}) \Delta Z/D_{MW}^{3} $ where $\Delta$Z is the projected distance on the line-of-sight axis (see Figure~\ref{FigS2}). The potential associated to the tidal force \citep{Souchay2013} exerted by the MW on the dSph at a projected radius R= $r_{half}$ is written as:

\begin{equation}
\phi_{tidal}  \sim - G M_{MW} \times \frac{Z^2}{D_{MW}^3}
\label{EqS6}
\end{equation}

One needs to calculate the average potential for a Plummer profile (Eq.~\ref{EqS1}), leading to:

\begin{equation}
<\phi_{tidal}> = - \frac{G M_{MW}}{D_{MW}^3} \times \frac{\int_{-\infty}^{+\infty} \rho(Z) Z^2 dZ}{\int_{-\infty}^{+\infty} \rho(Z) dZ }  
\label{EqS7}
\end{equation}

And after integration, this leads to:

\begin{equation}
<\phi_{tidal}> = \frac{G M_{MW}}{D_{MW}^3} \times r_{half}^2   
\label{EqS8}
\end{equation}

If we consider a system for which kinematics are driven by Galactic tidal forces, one would have $<\phi_{tidal}>$ + K = 0 (energy conservation), and then:

\begin{equation}
\sigma_{los,tidal}^2 = \frac{ 2 G M_{MW}(D_{MW}) \times r_{half}^2 }{D_{MW}^3}  
\label{EqS9}
\end{equation}

In such a case the predicted total mass would be: 

\begin{equation}
M_{los,tides pred} = 4.989677 G M_{MW} \times \frac{ r_{half}^3}{D_{MW}^3}   
\label{EqS10}
\end{equation}

\begin{figure}
\epsscale{0.9}
\plotone{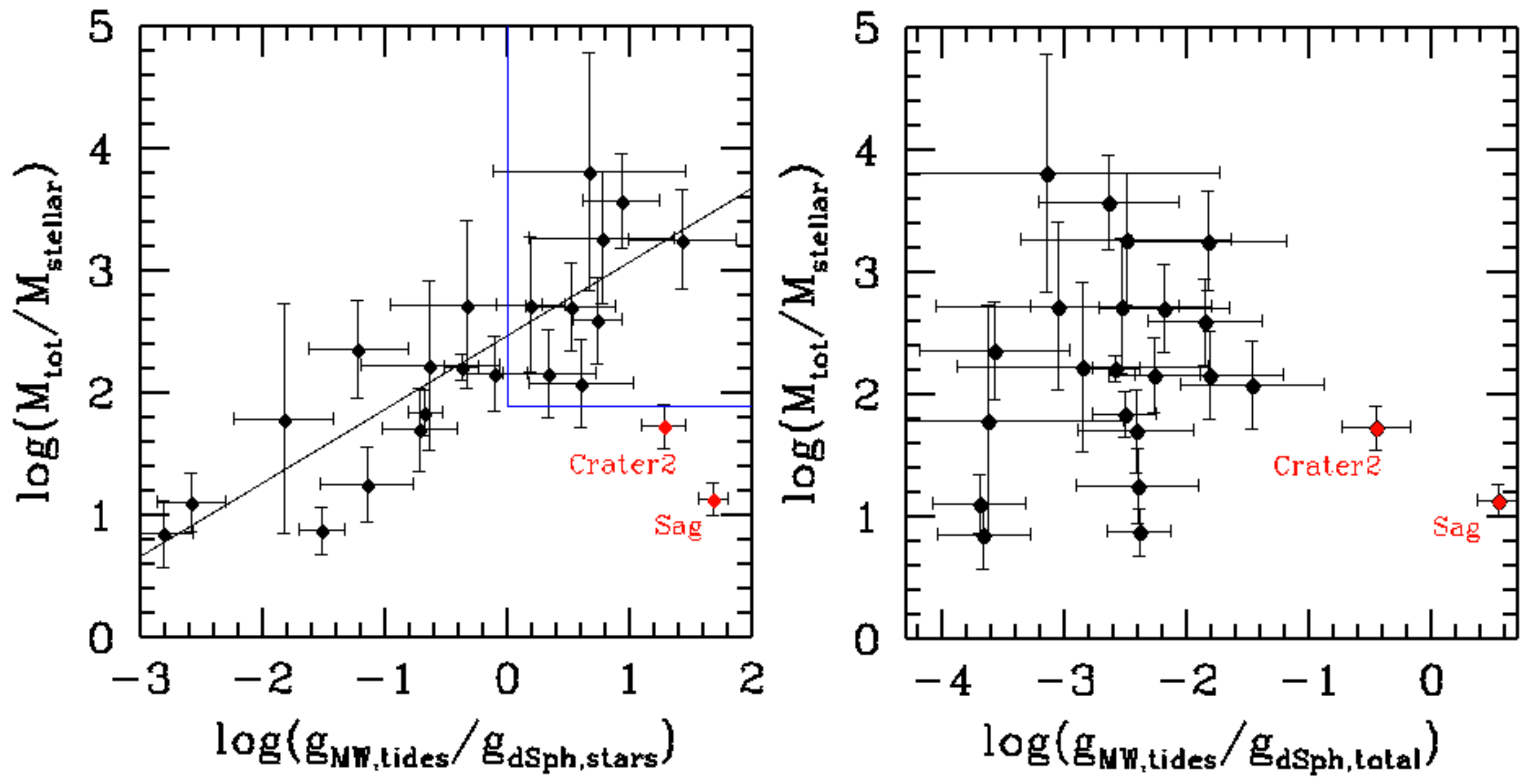}
\caption{{\it(Left)}: total-to-stellar mass ratio as a function of the ratio between the MW tides and the dSph stellar content acceleration. Sagittarius and Crater 2 are indicated as outliers of the correlation with $\rho$= 0.82, t= 6.3 and P= 1.9 $10^{-6}$. The blue box isolates eight ultra-faint dwarfs and Sextans (see the text). {\it(Right)}: the same, but replacing $M_{stellar}$ by $M_{tot}$ in the calculation of the dSph self-gravity acceleration. The correlation vanishes ($\rho$= 0.27, t= 1.24 and P= 0.11), which evidences further that total mass estimates based on an equilibrium assumption for dSphs depend indeed on the MW forces that are dissolving them.
}
\label{FigS3}
\end{figure}

This also results in a $g_{MW,tides}/g_{dSph,stars}$ ratio that is precisely the ratio between the above predicted mass and half the stellar mass. Figure~\ref{FigS3} shows the correlation between total-to-stellar mass and $g_{MW,tides}/g_{dSph,stars}$  ratios. However it is far less significant than that with Galactic acceleration ($g_{MW}$), as shown in the top-left panel of Figure~\ref{Fig2}. As in Figure~\ref{Fig2}, the correlation disappears when replacing $gd_{Sph,stars}$ by $g_{dSph,tot}$, which further supports the intimate link between total mass estimate from kinematics and Galactic gravitational forces.\\

The left panel of Figure~\ref{FigS3} shows that at R= $r_{half}$, the tidal acceleration exerted by the MW dominates that of the dSph stellar content for 8 Ultra Faint dwarfs (UFDs, Segue I, UrsaMajor 2, Bootes 2, Segue 2, Coma Berenices, Bootes, Ursa Major, Hercules) and Sextans. Most of these objects have their major axes aligned with the VPOS suggesting further that tidal effects are at work. We also note that Hercules is likely in tidal disruption \citep{Garling2018}, while its  location and elongation (see Figure~\ref{Fig1}) suggest an orbit close to that of Sagittarius.  \\

Figure~\ref{Fig4} shows that the tidal forces exerted by the MW are likely controlling the dynamics of Sagittarius. The large value of the acceleration ratio in Figure~\ref{FigS3} confirms this. In Figure~\ref{Fig4}, the eight UFDs and Sextans are mostly below the dashed (best-fit) line suggesting that those objects may be transitioning toward tidally dominated galaxies.

\section{Morphologies and velocity gradients for simulated dwarfs}
\label{simus}
Simulations by \citet{Yang2014} used gas-rich objects to test their physical transformation (gas removal and tidal stirring) during their first passage into the MW halo (see the online animated Figure~\ref{fig5anim}). These are based on the suggestion \citep{Kroupa1997,Dabringhauser2013,Hammer2013,Yang2014} that MW dSphs might be residues of ancient, gas-rich, tidal dwarf galaxies (TDGs) to explain the VPOS thickness \citep{Pawlowski2014}. Observations of high-velocity clouds in the Magellanic Stream \citep{Kalberla2006} have shown that the MW is surrounded by a huge halo of low-density (i.e, 1 to 2. $10^{-4} atom.cm^{-3}$ at about 50 kpc) hot gas (T $\ge$ $10^6$ K) out to at least 300 kpc. Halo hot gas may induce ram pressure and hence could also play an important role in transforming gas-rich objects into dSphs \citep{Mayer2001}. Recall that in the $\Lambda$CDM framework, TDGs are devoid of dark-matter. \\ 

\begin{figure}
\epsscale{0.9}
\plotone{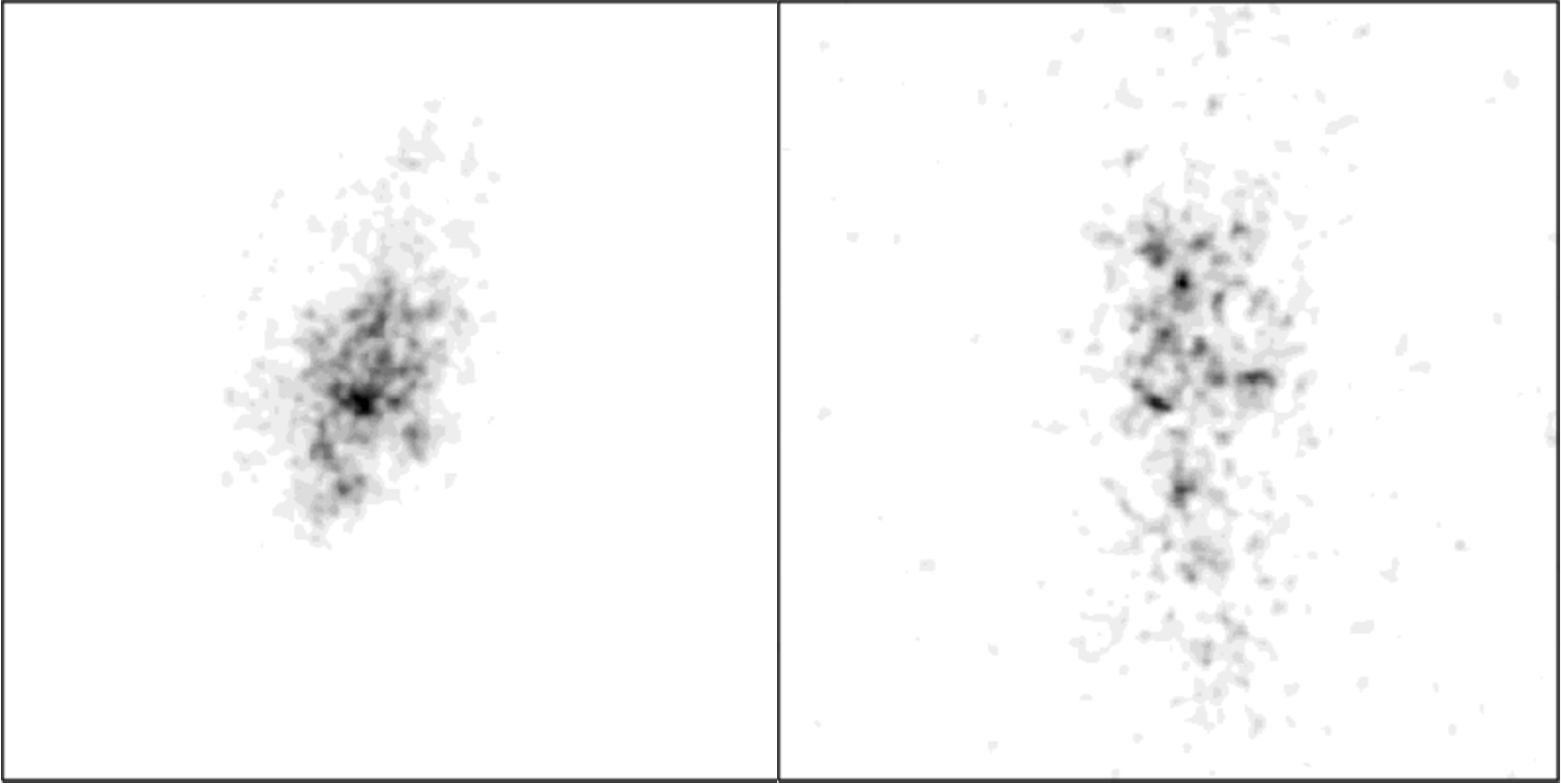}
\caption{Simulated images of the residuals of model TGD3-rp80 before (left, $T_{peri}$ = -0.11 Gyr, see Table~\ref{Tab3}) and after its passage at the pericenter (right, $T_{peri}$ =+0.46 Gyr, see Table~\ref{Tab3}). They show the elongation of the object along its trajectory, which is vertical here, since the images are shown in Galactic coordinates (l, b). The physical size of the image is 6x6 $kpc^2$, $L_V$ luminosity was approximated by $M_{\odot}/L_{\odot}$=1, and an artificial noise of 28 $mag.arcsec^{-2}$ (1$\sigma$) was added to account for the sky residuals.
}
\label{FigS4}
\end{figure}

\begin{figure}
\epsscale{0.9}
\plotone{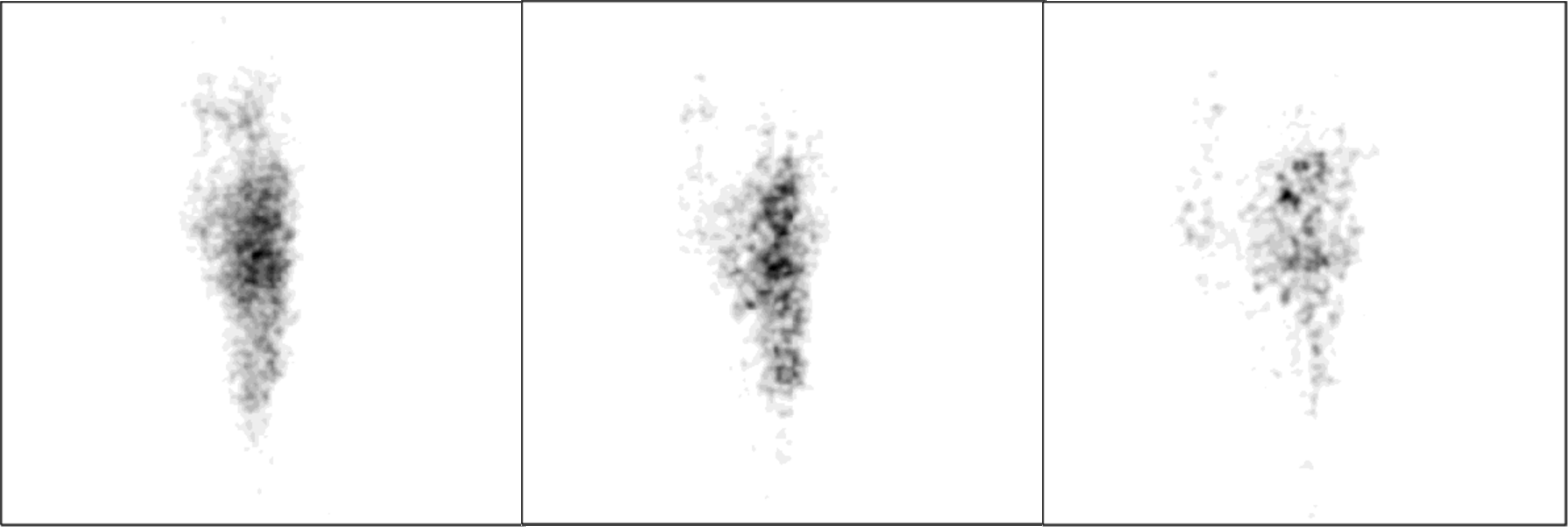}
\caption{Three snapshots images from model TGD3-rp16; from left to right, with $T_{peri}$= +0.18, +0.2 and +0.22 Gyr (see Table~\ref{Tab3}) after pericenter, respectively. They show a possible dislocation, perpendicular to the orbital motion along the vertical axis (b), which may occur when the pericenter is relatively small (25.1 kpc, see Table~\ref{Tab3}). These images were created using the same conditions than in Figure~\ref{FigS4} (physical size of 6x6 $kpc^2$).
}
\label{FigS5}
\end{figure}

These simulations \citep{Yang2014} are based on the hydrodynamical/N-body code Gadget2 \citep{Springel2005}, and are using simulated DM-free galaxies with initial mass and gas fraction (within a 1 kpc projected radius) of 1.35 $10^8 M_{\odot}$ and 71\%, respectively. For each realization, the stellar component is sampled by 500,000 particles. DM-free galaxies are then sent into the MW halo on hyperbolic orbits (i.e., first infall) with two different pericenters, which sample the range of MW dSphs orbits.  Table~\ref{Tab3} shows five different snapshots (epochs), the first two associated to a large pericenter, and the last three snapshots are describing the disruptive effect of passing at a small pericenter. Figures ~\ref{FigS4} and~\ref{FigS5} show that simulated galaxies are elongated along their trajectories in agreement with observations of most dSphs (see Figure~\ref{Fig1}). 

To estimate the velocity gradients of simulated galaxies and their preferential orientation $\theta$, the same procedure as for the observations was applied. For this we extracted a random subsample of the simulation, corresponding to the number of members in the observations and to which we added uncertainties of 2 km/s. Table~\ref{Tab3} gives the same quantities as Table~\ref{Tab2}, together with data relative to the simulated orbital motions and initial conditions. The simulations show similar velocity gradients and orientations as the observed galaxies. Figure~\ref{FigS5} captures a possible disruption of a dwarf probably caused by the axisymmetric action of the MW disk for a small pericenter (25.1 kpc) passage. This could correspond to Ursa Minor, which shows multiple components \citep{Bellazzini2002,Pace2014}, as well as a peculiar orientation of its velocity gradient (compare $\theta$ in Table~\ref{Tab2} with that in Table~\ref{Tab3} for TDG3-rp16 model at $T_{peri}$= +0.18).

%\begin{longrotatetable}
\begin{deluxetable*}{ccccccccccc}
%\tablenum{3}
\tablecaption{Velocity gradients and orbital parameters for modelled dSphs\label{Tab3}}
\tablewidth{100pt}
\tabletypesize{\scriptsize}
\tablehead{
\colhead{Model} & \colhead{$r_{half}$} & \colhead{N} & \colhead{$\Delta$X} & \colhead{$\theta$} & \colhead{P-value} & \colhead{$M_{stellar}$} & \colhead{$\sigma_{los}$} & \colhead{$D_{MW}$} & \colhead{$T_{peri}$} & \colhead{$R_{peri}$}\\
\colhead{name} & \colhead{(kpc)} & \colhead{particles} & \colhead{(kpc)} & \colhead{(degrees)} & \colhead{} & \colhead{$10^6$$M_{\odot}$} & \colhead{$km s^{-1}$} & \colhead{(kpc)} & \colhead{Gyr} & \colhead{(kpc)}
}
\decimalcolnumbers
\startdata
TDG3-rp80 &  0.68 &  2520 &  2.56 &  91 &  $<$0.0000001 &  0.96 &  3.7 &  90.5 &  -0.11 &  89.5\\
TDG3-rp80 &  0.70 &  1000 &  4.5 &  102 &  $<$0.0000001 &  0.63 &  5.3 &  142.9 &  +0.46 &  89.5\\
TDG3-rp16 &  1.29 &  500 &  2.42 &  5 &   0.000030 &  4.59 &  4.0 &  87.1 &  +0.18 &  25.1\\
TDG3-rp16 &  1.21 &  500 &  5.89 &  98 &   0.000001 &  4.08 &  4.5 &  96.1 &  +0.20 &  25.1\\
TDG3-rp16 &  1.13 &  500 &  4.64 &  105 &   0.000002 &  3.26 &  4.6 &  105. &  +0.22 &  25.1\\
\enddata
\tablecomments{This table provides parameters calculated for two simulations of a DM-free galaxy, so-called TDG3 \citep{Yang2014}, interacting with the MW halo and disk.  Half-mass radii of simulated objects are evaluated with a detection limit of 28 $mag.arcsec^{-2}$ in the V-band and stellar mass-to-light ratio of 1 in the V-band. $\Delta$X, $\theta$, and P-value have the same meaning as in Table 2. $T_{peri}$ provides the time before (minus values) or after (plus values) the passage at pericenter, $R_{peri}$.}
\end{deluxetable*}
%\end{longrotatetable}

There are, however, differences between simulations and observations that could be due to the impact of gas stripping, which affects more simulated galaxies than real dSphs. In the latter objects, the kinematics seem to be actually driven by Galactic acceleration. If gas stripping really occurred, it should have happened a long time before the passage at pericenter, which would require larger amounts of MW halo hot gas than what was adopted by \citet{Yang2014}.



\end{document}